\documentclass{article}
\usepackage{arxiv}
\usepackage[utf8]{inputenc} 
\usepackage[T1]{fontenc}    
\usepackage{booktabs}       
\usepackage{amsfonts,amsmath,amssymb}       
\usepackage{nicefrac}       
\usepackage{microtype}      
\usepackage{graphicx}
\usepackage[square,sort,comma,numbers]{natbib}
\usepackage{doi}
\usepackage[toc,page]{appendix}
\usepackage{url,hyperref,lineno,microtype,subcaption}
\usepackage[onehalfspacing]{setspace}
\usepackage{algorithmic}
\usepackage{booktabs}
\usepackage{array}
\usepackage{soul,color}
\usepackage{mathtools, esdiff}
\usepackage{geometry}
\usepackage[dvipsnames]{xcolor}

\newcommand{\Un}{U_\mathrm{n}}
\newcommand{\Up}{U_\mathrm{p}}
\newcommand{\Ui}{U_i}

\newcommand{\nmcsix}{\mathrm{Ni}_\mathrm{0.6}\mathrm{Mn}_\mathrm{0.2}\mathrm{Co}_\mathrm{0.2}}
\newcommand{\degC}{^\circ\mathrm{C}}

\newcommand{\Vt}{V_{\mathrm{t}}}
\newcommand{\Qp}{Q_{\mathrm{p}}}
\newcommand{\Qn}{Q_{\mathrm{n}}}
\newcommand{\Qsei}{Q_{\mathrm{SEI}}}
\newcommand{\Qseir}{Q_{\mathrm{SEI},r}}
\newcommand{\Qi}{Q_i}
\newcommand{\Bu}{\mathrm{Bu}}
\newcommand{\jsei}{j_{\mathrm{SEI}}}
\newcommand{\rhosei}{\rho_{\mathrm{SEI}}}
\newcommand{\msei}{m_{\mathrm{SEI}}}
\newcommand{\jseir}{j_{\mathrm{SEI},r}}
\newcommand{\Usei}{U_{\mathrm{SEI}}}
\newcommand{\Useir}{U_{\mathrm{SEI},r}}
\newcommand{\Isei}{I_{\mathrm{SEI}}}

\newcommand{\Rsei}{R_{\mathrm{SEI}}}

\newcommand{\Dsei}{D_{\mathrm{SEI}}}
\newcommand{\Dseirl}{D_{\mathrm{SEI},rl}}
\newcommand{\Dseirbar}{\bar{D}_{{\mathrm{SEI},r}}}
\newcommand{\DseirA}{D_{\mathrm{SEI},rA'}}
\newcommand{\DseirB}{D_{\mathrm{SEI},rB'}}
\newcommand{\ksei}{k_{\mathrm{SEI}}}
\newcommand{\kseir}{k_{\mathrm{SEI},r}}
\newcommand{\krxn}{k_{\mathrm{rxn}}}
\newcommand{\kdif}{k_{\mathrm{dif}}}
\newcommand{\vmsei}{V_{m,\mathrm{SEI}}}
\newcommand{\vsei}{V_{\mathrm{SEI}}}
\newcommand{\vmseir}{V_{m,\mathrm{SEI},r}}

\newcommand{\jseidif}{\tilde{j}_{\mathrm{SEI,dif}}}
\newcommand{\jseidifr}{\tilde{j}_{\mathrm{SEI,dif,r}}}
\newcommand{\jseirxn}{\tilde{j}_{\mathrm{SEI,rxn}}}
\newcommand{\jseirxnr}{\tilde{j}_{\mathrm{SEI,rxn,r}}}
\newcommand{\nup}{\nu_{\mathrm{p}}}
\newcommand{\nun}{\nu_{\mathrm{n}}}

\newcommand{\Iint}{I_{\mathrm{int}}}
\newcommand{\Iapp}{I_{\mathrm{app}}}
\newcommand{\japp}{j_{\mathrm{app}}}
\newcommand{\Msei}{M_{\mathrm{SEI}}}

\newcommand{\phisn}{\phi_{\mathrm{s,n}}}

\newcommand{\phie}{\phi_{\mathrm{e}}}
\newcommand{\thetap}{\theta_{\mathrm{p}}}
\newcommand{\thetaps}{\theta_{\mathrm{p}}^\mathrm{s}}
\newcommand{\thetan}{\theta_{\mathrm{n}}}
\newcommand{\thetans}{\theta_{\mathrm{n}}^\mathrm{s}}
\newcommand{\thetai}{\theta_i}

\newcommand{\An}{A_{\mathrm{n}}}

\newcommand{\Ln}{L_{\mathrm{n}}}
\newcommand{\Lp}{L_{\mathrm{p}}}
\newcommand{\Li}{L_{\mathrm{i}}}
\newcommand{\Rn}{R_{\mathrm{n}}}

\newcommand{\Ri}{R_{\mathrm{i}}}

\newcommand{\etasei}{\eta_{\mathrm{SEI}}}
\newcommand{\etai}{\eta_i}
\newcommand{\etan}{\eta_\mathrm{n}}
\newcommand{\etap}{\eta_\mathrm{p}}
\newcommand{\asn}{a_{\mathrm{s,n}}}
\newcommand{\alphasei}{\alpha_{\mathrm{SEI}}}
\newcommand{\alphaseir}{\alpha_{\mathrm{SEI},r}}
\newcommand{\cecs}{c_{\mathrm{EC}}^{\mathrm{s}}}
\newcommand{\ceco}{c_{\mathrm{EC}}^0}
\newcommand{\cseir}{c_{\mathrm{SEI},r}}
\newcommand{\cseio}{c_{\mathrm{SEI},r}^0}

\newcommand{\csps}{c_{\mathrm{s,p}}^{\mathrm{s}}}
\newcommand{\cspm}{c_{\mathrm{s,p}}^{\mathrm{max}}}

\newcommand{\csns}{c_{\mathrm{s,n}}^{\mathrm{s}}}
\newcommand{\csnm}{c_{\mathrm{s,n}}^{\mathrm{max}}}

\newcommand{\csis}{c_{\mathrm{s,i}}^{\mathrm{s}}}
\newcommand{\csia}{c_{\mathrm{s,i}}^{\mathrm{avg}}}
\newcommand{\deltasei}{\delta_{\mathrm{SEI}}}
\newcommand{\deltaseir}{\delta_{\mathrm{SEI},r}}
\newcommand{\deltaseibar}{\bar{\delta}_{\mathrm{SEI}}}

\newcommand{\Ron}{R_{\mathrm{ct,n}}}
\newcommand{\Roi}{R_{\mathrm{ct},i}}

\newcommand{\Rwi}{R_{\mathrm{diff},i}}

\newcommand{\Cwi}{C_{\mathrm{diff},i}}

\newcommand{\taui}{\tau_{\mathrm{diff},i}}

\newcommand{\Iwi}{I_{\mathrm{diff},i}}
\newcommand{\liedc}{\mathrm{Li_2EDC}}

\newcommand{\ec}{\mathrm{EC}}
\newcommand{\lip}{\mathrm{Li^+}}
\newcommand{\electron}{\mathrm{e^-}}
\newcommand{\ethylene}{\mathrm{C_2H_4}}
\definecolor{lightblue}{rgb}{0.7,1,1}
\definecolor{lightgray}{rgb}{0.9,0.9,0.9}
\definecolor{lightgreen}{rgb}{0.7,1,0.7}

\newif\ifrespondingtoreviewers
\respondingtoreviewersfalse

\ifrespondingtoreviewers
    \newcommand{\reva}[1]{\sethlcolor{yellow}\hl{#1}}

    \newcommand{\rev}[1]{\sethlcolor{lightgray}\hl{#1}}
\else
    \newcommand{\rev}[1]{#1}
    \newcommand{\reva}[1]{#1}

\fi

\newcommand{\mytitle}{Modeling Battery Formation: Boosted SEI Growth, Multi-Species Reactions, and Irreversible Expansion}
\renewcommand{\headeright}{}
\renewcommand{\undertitle}{}
\renewcommand{\shorttitle}{Modeling Battery Formation}

\title{\mytitle}

\author{
        Andrew Weng \\
	Mechanical Engineering\\
	University of Michigan\\
	Ann Arbor, MI 48109 \\
	\texttt{asweng@umich.edu} \\
    \And
	Everardo Olide \\
	Applied Physics \\
	University of Michigan\\
	Ann Arbor, MI 48109\\
	\texttt{eolide@umich.edu} \\
   \And 
        Iaroslav Kovalchuk \\
        Electrical Engineering and Computer Science\\
        University of Michigan\\
        Ann Arbor, MI 48109\\
        \texttt{ikovalch@umich.edu} \\
    \And 
	Jason B.~Siegel \\
	Mechanical Engineering\\
	University of Michigan\\
	Ann Arbor, MI 48109\\
	\texttt{siegeljb@umich.edu} \\
    \And
    Anna Stefanopoulou \\
    Mechanical Engineering \\
    University of Michigan \\
    Ann Arbor, MI 48109 \\
    \texttt{annastef@umich.edu}
}

\hypersetup{
pdftitle={\mytitle},
pdfsubject={q-bio.NC, q-bio.QM},
pdfauthor={Andrew Weng, Everardo E.~Olide, Jason B.~Siegel, Anna Stefanopoulou},
pdfkeywords={First keyword, Second keyword, More},
}

\begin{document}
\maketitle

\begin{abstract}
This work proposes a semi-empirical model for the SEI growth process during the early stages of lithium-ion battery formation cycling and aging. By combining a full-cell model which tracks half-cell equilibrium potentials, a zero-dimensional model of SEI growth kinetics, and a semi-empirical description of cell thickness expansion, the resulting model replicated experimental trends measured on a 2.5 Ah pouch cell, including the calculated first-cycle efficiency, measured cell thickness changes, and electrolyte reduction peaks during the first charge dQ/dV signal. This work also introduces an SEI growth boosting formalism that enables a unified description of SEI growth during both cycling and aging. This feature can enable future applications for modeling path-dependent aging over a cell's life. The model further provides a homogenized representation of multiple SEI reactions enabling the study of both solvent and additive consumption during formation. This work bridges the gap between electrochemical descriptions of SEI growth and applications towards improving industrial battery manufacturing process control where battery formation is an essential but time-consuming final step. We envision that the formation model can be used to predict the impact of formation protocols and electrolyte systems on SEI passivation and resulting battery lifetime. 
\end{abstract}


\newpage

\section{Introduction}

Every commercial lithium-ion battery undergoes formation cycling and aging at the end of the battery cell manufacturing process \citep{An2016-kk, Liu2021-ye}. The formation process is time and capital-intensive, motivating battery manufacturers to develop new formation protocols to decrease formation time while maintaining battery lifetime and safety \citep{Liu2021-ye}. Yet, despite the importance of battery formation, a general framework for modeling the formation process for commercial lithium-ion battery systems is lacking. \rev{Without these models, formation protocol optimization will require brute-force, trial-and-error approaches, which are slow, inefficient, and not guaranteed to yield optimal outcomes.}

The main goal of battery formation is to form a passivating solid electrolyte interphase (SEI) layer at the negative electrode surface which limits further SEI growth over the battery's life \citep{Balbuena2004-dy, Pinson2012-wt, Peled2017-ob, Heiskanen2019-an}. SEI growth occurs throughout the battery formation process which consists of both cycling and a calendar aging steps. During formation cycling, the battery is externally charged and discharged for the first time following electrolyte filling \cite{Liu2021-ye}. Formation cycling is followed by formation aging, during which the battery cells are stored at high temperatures and high states of charge (SOCs) for days to weeks to continue the SEI growth process and screen for quality defects \citep{Liu2021-ye}. 

The SEI reaction and film growth process is complex. Multiple reaction pathways are often involved since multiple electrolyte components, including solvents and additives, can participate simultaneously in the SEI reaction \citep{An2016-kk, Petibon2016-rc, Peled2017-ob}. The resulting SEI film is thus heterogeneous in composition \citep{Roder2017-ox, Huang2019-qh, Peled2017-ob}. The SEI film is also difficult to study experimentally owing to the reactivity of the electrode-electrolyte interface and the nanometer thickness of the film \citep{An2016-kk, Peled2017-ob}. \rev{These inherent complexities of the SEI growth process partly explain why existing SEI growth models are difficult to parameterize and experimentally validate, hindering their applications in a battery manufacturing context for formation process design and lifetime prediction.}

\subsection{A Phenomenological Basis for Formation Modeling}

\begin{figure}[ht!]
\begin{center}
\includegraphics[width=\linewidth]{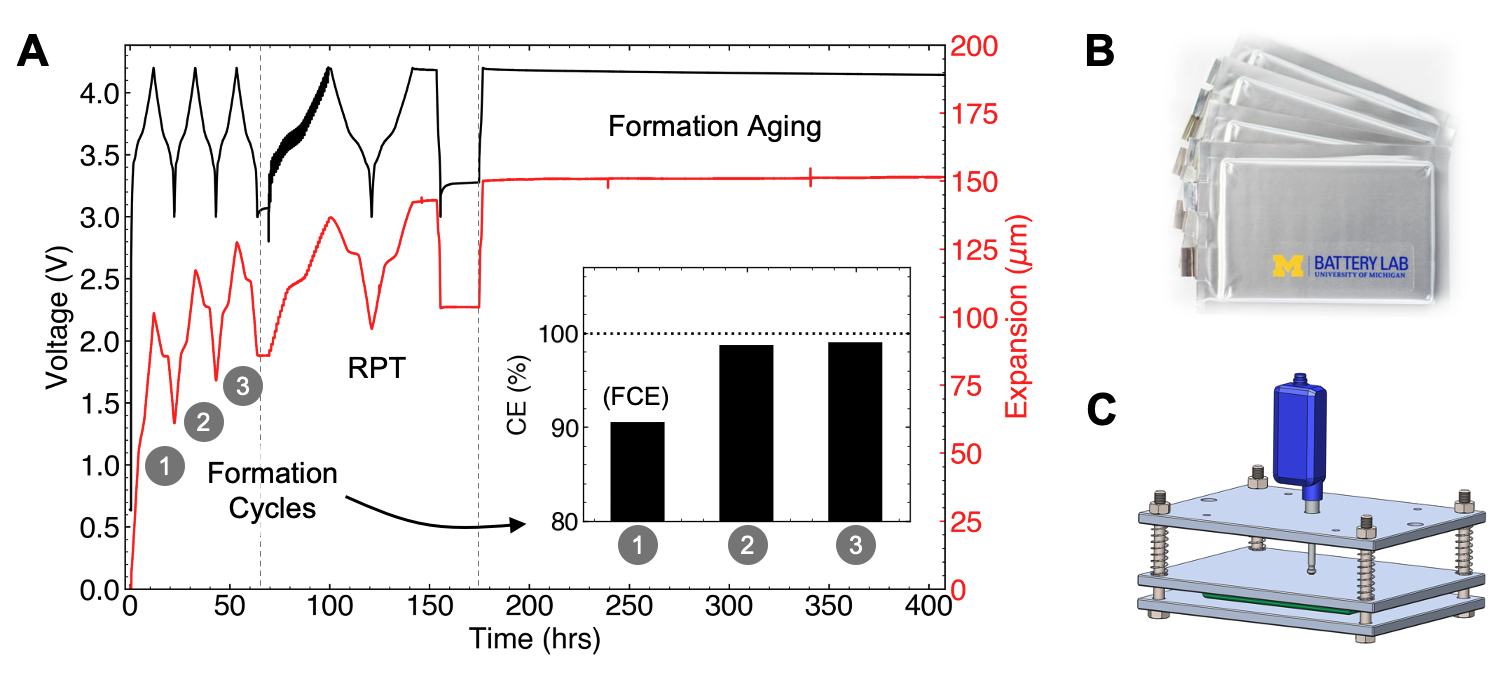}
\end{center}
\caption{\textbf{Experimentally-measurable macroscopic signals from the battery formation and aging process.} (A) Electrochemical (voltage and current) signal is shown in black. \rev{These signals are obtainable directly from the equipment used for battery formation cycling.} Macroscopic cell thickness expansion is shown in red. (B) Example image of the multi-layer stacked pouch cells similar to the one used for this work. (C) An expansion fixture instrumented with a linear displacement sensor enabling real-time cell thickness expansion during the formation process.}
\label{fig:experimental}
\end{figure}

Despite the microscopic complexities of SEI growth, the battery formation process for commercial-scale cells \rev{yields} electrochemical signals which can be directly measured \rev{using standard equipment during formation cycling}. Figure \ref{fig:experimental} shows example data collected on a 2.5 Ah pouch cell (Panel B) that underwent three formation charge-discharge cycles, followed by reference performance tests (RPTs), followed by high-temperature formation aging (see Section \ref{sec:cellbuild} for more experimental details). This dataset can be directly used to calculate the coulombic efficiency (CE), according to:
\begin{equation} \label{eq:ce}
    \mathrm{CE}_i = \frac{Q_{\mathrm{d},i}}{Q_{\mathrm{c},i}},
\end{equation}
where ${Q_{\mathrm{d},i}}$ and $Q_{\mathrm{c},i}$ are the discharge and charge capacities of the $i$th cycle, calculated via current integration. The CE of the first cycle, also known as the first cycle efficiency (FCE), is characteristically lower than the CE for all subsequent cycles owing to the rapid consumption of lithium during the first charge cycle to form the SEI. 

\rev{For pouch cell form factors, thickness expansion can also be directly measured} using a sensor fixture shown in Figure \ref{fig:experimental}C \citep{Mohtat2020-zp, Mohtat2021-qh, Pannala2022-cz}. The measurements suggest that total cell expansion can be attributed to two distinct sources. The first source is due to changes in the lithium content (or stoichiometry) in the positive and negative electrode particles. During lithiation, the particles swell, and during delithiation, the particles contract. Since this process is reversible, we will refer to this source as ``reversible expansion.'' The reversible expansion tracks closely to the measured cell voltage, or state of charge (SOC), which determines the lithium content in either electrode. The second source of expansion can be noticed from the  \textit{minimum expansion} at the end of every discharge cycle. This expansion appears to always increase over cycles. We will thus refer to this expansion source as ``irreversible expansion'' \citep{Louli2019-wc, Mohtat2021-qh}. The irreversible expansion is attributed to the growth of the SEI layer during formation and aging. Consistent with the CE data, the irreversible expansion is high during the first cycle, but then slows down over the next two cycles. During formation aging, the irreversible expansion rate appears to decrease, indicating a possibly slower SEI growth process compared to cycling.

\subsection{Main Contributions}

This work presents a semi-empirical modeling framework describing the SEI growth process during battery formation cycling and aging. The model builds on established electrochemical descriptions of SEI growth which uses Butler-Volmer kinetics to describe the SEI-forming solvent reduction process \citep{Ramadass2004-oz, Ning2006-hb, Safari2008-zk, Ekstrom2015-yz} and linearized Fick's law to describe solvent diffusion \citep{Ploehn2004-fq, Reniers2019-rp, Pinson2012-wt, Kamyab2019-uq}. To model cell expansion, a semi-empirical approach is taken which draws from existing literature \citep{Kupper2018-qh, Laresgoiti2015-qj, Mohtat2020-zp, Brosa_Planella2022-aq, Wang2022-tl, OKane2022-fj}, but separates the reversible and irreversible expansion contributions to the total expansion. Since our aim is to develop a reduced-order description of SEI growth which can be deployed in a battery manufacturing context, our modeling approach does not explicitly consider spatial variations in the SEI layer \cite{Christensen2004-im, Single2016-bd, Single2017-ei, Von_Kolzenberg2022-sd} or competing reaction pathways \cite{Roder2017-ox}. However, as will be demonstrated, a reduced-order, zero-dimensional model can accurately capture macroscopic, observed trends during both formation cycling and aging shown in Figure \ref{fig:experimental}, including CE, expansion, and first cycle dQ/dV peaks indicating electrolyte reduction processes.

Our model introduces three extensions to the existing literature on zero-dimensional SEI growth models. First, we introduce a semi-empirical model of macroscopic thickness expansion of the battery, accounting for both the reversible expansion due to lithium intercalation reactions as well as the irreversible expansion due to SEI growth (see Section \ref{sec:expansion}). Second, we introduce a mathematical formalism to describe ``boosted SEI growth'' during charging (see Section \ref{sec:boost}). This model extension unifies the description of SEI growth and cell expansion during cycling, which is fast, and during calendar aging, which is slow. Finally, we introduce a \rev{homogenized, multi-species} representation of the SEI growth process, enabling the prediction of volume-averaged SEI film properties under multiple reacting species (see Section \ref{sec:homogenized}). This model extension enables a description of solvent consumption and additive consumption as two parallel and coupled processes which combine to create a composite SEI film \rev{with a volume-averaged film diffusivity}.

The formation model we develop resolves the dynamics of lithium consumption and cell expansion during formation cycling and aging. \rev{The same model can be applied} to track lithium inventory loss over the remaining life of the cell, including during cyclic aging and calendar aging. By explicitly considering multiple electrolyte reduction reactions during the first cycle, the model also provides a pathway for future studies on how formation protocols influence SEI passivation properties \cite{Attia2021-bl} and their consequences on battery lifetime.

\section{Experimental Methods}

Figure \ref{fig:experimental} summarizes the experimental data collected for this work. This dataset defines a basic set of macroscopic observations that the formation model will seek to capture. The experiment consists of three parts: (1) formation cycling, (2) reference performance testing (RPTs), and (3) formation aging at 100\% state of charge (SOC). The proceeding sections describe the cell build process in relation to the formation cycling and aging experiments shown in Figure \ref{fig:experimental}A.

\subsection{Cell build} \label{sec:cellbuild}

Experimental data was collected on a 2.5Ah, multi-layer stacked pouch cell (Figure \ref{fig:experimental}B). The cell was built on a prototype cell manufacturing line using a process similar to the one outlined in Weng et al. \citep{Weng2021-qc}. The cell consisted of a single-crystal $\nmcsix$ (NMC622) positive electrode (Targray) and a graphite negative electrode (Superior SLC, 1520-T), based on the available stock. A standard electrolyte formulation was used (SoulBrain PuriEL R\&D 326), which consisted of ethylene carbonate (EC) and ethyl methyl carbonate (EMC) in a 3:7 weight ratio, 1.0M LiPF$_6$ and 2 wt\% vinylene carbonate (VC) additive. The cell stack consisted of 7 double-sided positive electrode layers and 8 double-sided negative electrode layers, resulting in 14 active unit cells along the pouch cell thickness direction. The positive electrode formulation consisted of NMC622:C65:PVDF (94:3:3), where C65 and PVDF are binder materials. The negative electrode consisted of graphite:CMC:SBR (97:1.5:1.5), where CMC and SBR are binder materials. Positive (negative) electrode single-sided loadings were 17.2 (9.7) mg/cm$^2$ and with porosity targets of 30\% (20\%). The negative-to-positive capacity ratio was targeted to be 1.1.

\subsection{Electrolyte filling and cell fixture}

After electrolyte filling and enclosure sealing, cells were loaded in a custom-built pressure fixture shown in Figure \ref{fig:experimental}C. The pressure fixture was developed based on work by Mohtat et al. \cite{Mohtat2020-zp}. The fixture applied 5 psi of compression via a spring-loaded plate. To minimize copper dissolution at high negative electrode potentials immediately after electrolyte filling \citep{Guo2021-fa}, the cell was tap charged to 2.8V at a C/3 charge rate. The cell was subsequently left overnight to ensure the electrodes were completely wetted before formation charging. 

\subsection{Formation cycling and aging} 

Formation cycling was conducted while the cell/fixture assembly was placed in a temperature-controlled oven set to 45$\degC$. The fixture maintained a pressure of 5 psi throughout formation cycling and aging. For formation cycling, the cell was charged and discharged using an Arbin BT2000 system. 

The formation protocol consisted of three back-to-back charge-discharge cycles at a rate of C/10. Each charge terminated when the cell reached 4.2V, followed by a CV hold step with a C/20 current cut-off. Each discharge terminated when the cell reached 3.0V without CV holding. No rest periods were included between each charge and discharge.

Formation cycling was followed by a reference performance test (RPT), which included a pulse charge-discharge sequence for calculating cell resistances and C/20 charge-discharge voltage curves for dV/dQ analysis \citep{Weng2021-qc}. The details and contents of the RPT are not immediately relevant to this work, but the RPT data has been presented here to preserve data continuity with the subsequent experimental step.

After the RPT, the cell was charged to 100\% SOC to undergo formation aging at 45$\degC$. The goal of this step was to ensure SEI passivation and to screen for quality defects \citep{Liu2021-ye}. This aging step lasted for 14 days. After aging, the cell was finally discharged. Note that Figure \ref{fig:experimental}A only shows the first 10 days of the 14-day aging protocol for brevity.

\subsection{Expansion measurements} 

Thickness expansion of the cell was measured using a linear displacement sensor (Keyence GT2 series) mounted to the fixture (Figure \ref{fig:experimental}C). Expansion data was logged in real-time using a data acquisition system (LabVIEW) throughout the entire experiment. The expansion data was combined with the electrical signals during post-processing.

\section{Model Formulation}

\subsection{Model Assumptions}

\begin{figure}[ht!]
\begin{center}
\includegraphics[width=1.0\linewidth]{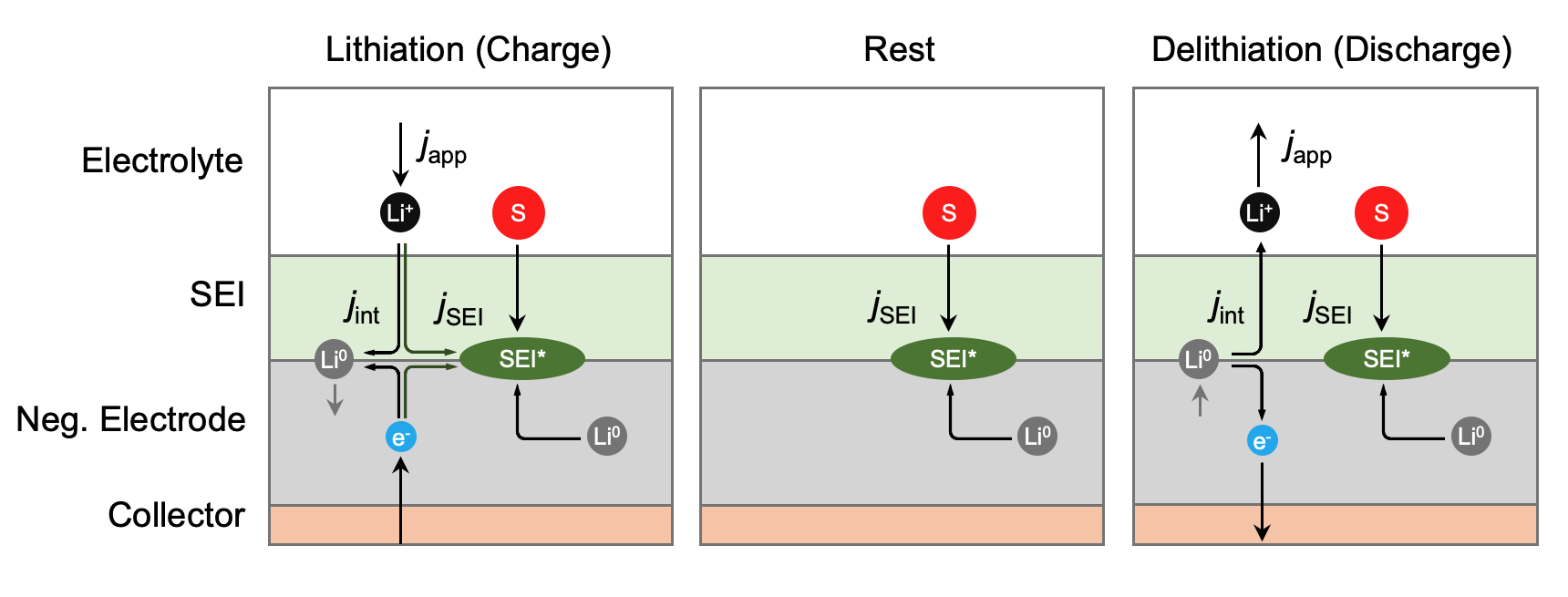}
\end{center}
\caption{\textbf{\rev{Consolidating SEI reaction pathways during charging, discharging and rest.}} S: solvent molecule, e.g. ethylene carbonate (EC) or vinylene carbonate (VC). SEI*: newly SEI formed at the electrode-SEI interface. Li$^+$: solvated lithium ions from the electrolyte. Li$^0$: neutral lithium intercalated into the negative electrode.}
\label{fig:schematic}
\end{figure}

Figure \ref{fig:schematic} shows the canonical reaction pathways for SEI formation \citep{Pinson2012-wt, Attia2020-hn} assumed for this work. The general SEI reaction is represented by: 
\begin{equation}
    \label{eq:sei-reaction}
    n\lip + n\electron + S \rightarrow \mathrm{SEI}\downarrow + P\uparrow,
\end{equation}
where $S$ represents a solvent molecule, $n$ is the number of participating electrons, `SEI' stands for the newly-formed solid reduction product, and $P$ denotes some reaction byproduct, usually a gas \citep{Rowden2020-yl}. Candidate solvent molecules include ethylene carbonate (EC) and  diethyl carbonate (DEC). Note that $S$ can also represent electrolyte additives such as vinylene carbonate (VC).

To build a model of SEI growth during both formation cycling and aging, SEI reaction pathways occurring during full cell charging, discharging, and rest, needed be consolidated. Doing so enables the same modeling framework to simulate cycling and calendar aging after formation completes. The reaction formulation we chose, summarized by Attia et al. \citep{Attia2020-hn}, considers two distinct SEI reaction modes, termed `electrochemical' SEI growth, which occurs in the presence of external current (i.e. during charge and discharge), and `chemical' SEI growth, which always occurs. Both SEI growth mechanisms share the same general reaction scheme described by Eq. \ref{eq:sei-reaction}, but only differ by the source of the lithium. In `electrochemical' SEI growth, the external current drives solvated lithium ions from the electrolyte towards the reaction interface. In `chemical' SEI growth, intercalated lithium from the negative electrode migrates to the reaction interface. These reaction modes will be unified by the model formulation presented in Section \ref{sec:sei}.

Further general modeling assumptions are listed below:

\begin{itemize}
    \item \textit{Reaction interface.} The SEI-forming reaction takes place exclusively at the electrode-SEI interphase \citep{Pinson2012-wt}.
    \item \textit{Rate-limiting mechanisms.} The SEI reaction rate is determined by two processes: (1) interfacial reaction kinetics according to Butler-Volmer kinetics, and (2) diffusion-limited solvent transport through a porous SEI. No other rate-limiting mechanisms such as electron conduction \citep{Single2016-bd} are considered.
    \item \textit{Idealized degradation.} SEI reaction at the negative electrode is the only source of full cell capacity loss. Considerations for other degradation modes, such as \reva{thermal SEI decomposition} \citep{Attia2020-hn}, lithium plating \citep{Ryan2021-ze}, active material degradation (e.g. cathode phase transformations \citep{Kang2008-aw}, particle cracking \citep{Smith2021-qv}, binder delamination \citep{Xu2018-xt}), and positive electrode side-reactions \citep{Xu2022-kg, Sungjemmenla2022-gm}, are left for future work.
    \item \textit{Irreversibility.} SEI reactions are irreversible.
    \item \textit{No gas formation.} Gas-forming reaction dynamics \citep{Rowden2020-yl} are ignored and left for future work.
    \item \textit{No cross-talk.} Cross-reactions between different electrolyte components is left as future model extensions.
\end{itemize}

\subsection{Electrode Potentials and Solid-Phase Lithium Stoichiometries}
\label{sec:eocp}

\begin{figure}[ht!]
\begin{center}
\includegraphics[width=0.85\linewidth]{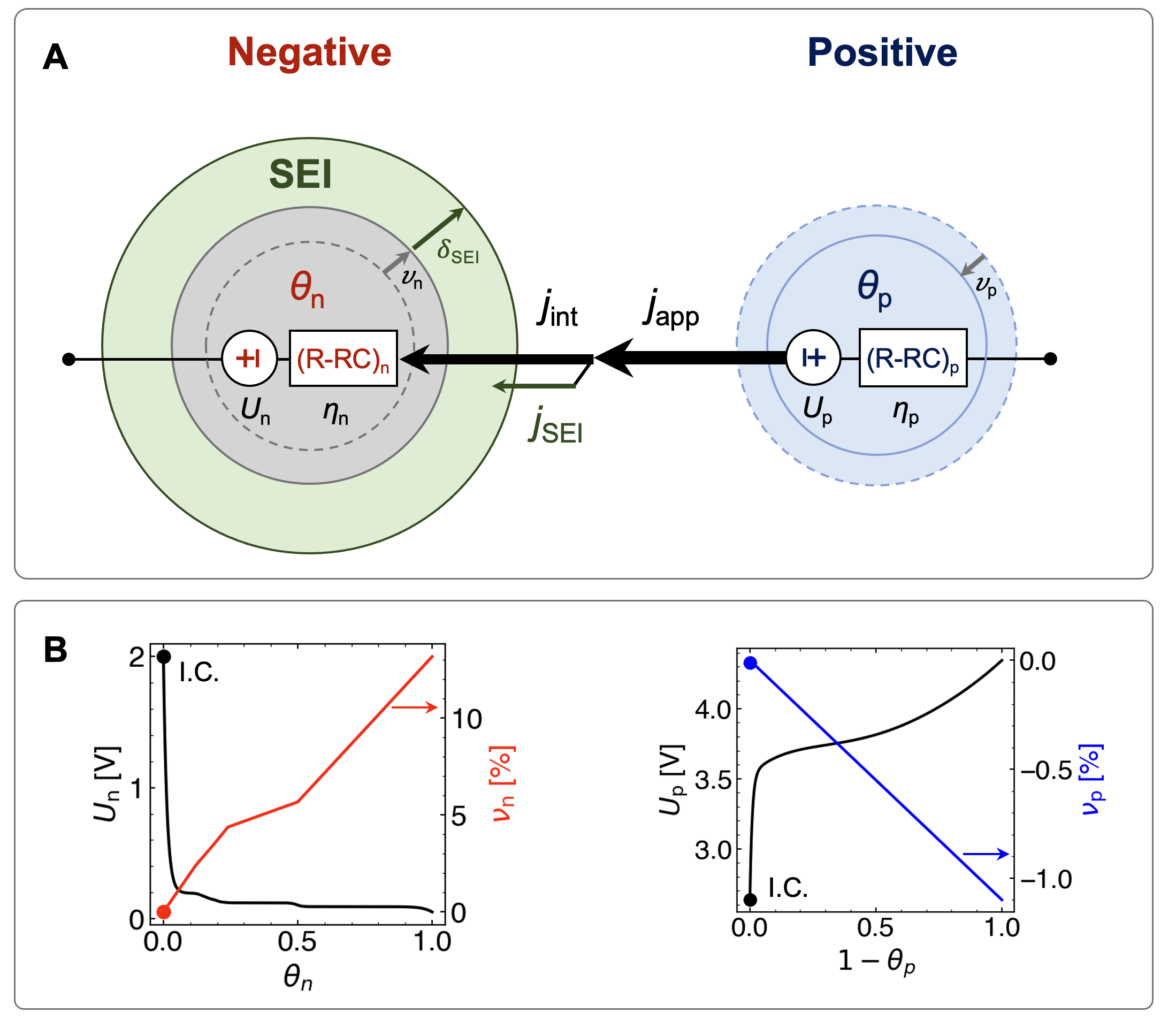}
\end{center}
\caption{\textbf{Zero-dimensional `dual-tank' formation modeling framework.} (A) Model components and key state variables. At each electrode $i=(\mathrm{n},\mathrm{p})$, the lithium stoichiometry $\thetai$ is tracked and used to update the electrode equilibrium surface potentials $U_\mathrm{i}$ and volumetric expansions $\nu_\mathrm{i}$. Electrode surface overpotentials are represented by R-RC elements which capture both charge-transfer and solid-state diffusion dynamics. The total applied current density $\japp$ passes through the positive electrode. However, at the negative electrode, the current is split between the intercalation current density $j_{\mathrm{int}}$ and the SEI current density $\jsei$. SEI build-up leads to SEI thickness growth, denoted by $\deltasei$. (B) Electrode equilibrium potential functions and volumetric expansion functions assumed for this work, adapted from Mohtat et al. \citep{Mohtat2020-zp}. The markers indicate the initial conditions (I.C.s) corresponding to the state of the cell before formation begins.}
\label{fig:model}
\end{figure}

A reduced-order full-cell model was used as a starting point for this work, shown in Figure \ref{fig:model}. In this model, the negative electrode equilibrium surface potential, $\Un$, forms the thermodynamic basis for SEI-forming reactions. However,  $\Un$ is not directly controllable or observable \rev{during the formation process for commercial devices. Rather, the terminal voltage,} $\Vt$, is observed and controlled. A practical model of battery formation therefore requires a description of $\Vt$, written as:
\begin{equation} \label{eq:vt}
    \Vt = \Up(\thetaps) - \Un(\thetans) + \etap + \etan,
\end{equation}
where $\Up$ is the positive electrode equilibrium surface potential, $\thetaps$ and $\thetans$ are the lithium stoichiometries at the electrode surfaces, and $\etap$ and $\etan$ are electrode overpotentials. $\Un$ and $\Up$ are \rev{described by empirical functions such as the ones shown in Figure} \ref{fig:model}B. The lithium stoichiometries are defined by:

\begin{align} \label{eq:thestoich}
    \thetaps &= \csps/\cspm \\
    \label{eq:thestoich2}
    \thetans &= \csns/\csnm,
\end{align}
where $\csps$ ($\csns$) is the lithium concentration at the surface of the positive (negative) electrode, and $\cspm$ ($\csnm$) is the maximum lithium concentration of the positive (negative) electrode. Finding the lithium surface concentrations will typically require solving the spherical diffusion equation \citep{Lin2011-mz}. Here, we simplify the representation by assuming that the current density is sufficiently low such that the solid-phase concentration gradient is approximately zero, hence $\csis \approx \csia$. Hence, the lithium stoichiometries can be directly updated via Coulomb counting the intercalation current:
\begin{align}
    \label{eq:dthetapdt}
    \frac{d\thetap(t)}{dt} &= -\frac{\Iapp(t)}{\Qp} \\ 
    \label{eq:dthetandt}
    \frac{d\thetan(t)}{dt} &= +\frac{\Iint(t)}{\Qn},
\end{align}
where $\thetap$ and $\thetan$ are the average lithium stoichiometries, $\Qp$ and $\Qn$ are the total electrode capacities corresponding to $\cspm$ and $\csnm$, respectively, $\Iapp$ is the applied current into the full cell, and $\Iint$ is the intercalation current at the negative electrode. We take the convention that $\Iapp>0$ when the cell is charged. Similarly, $\Iint > 0$ corresponds to lithium intercalation into the negative electrode. At the positive electrode, the intercalation current and the applied current are equal since our model assumed no side reactions at the cathode. At the negative electrode, the applied current is split between the intercalation current and the SEI reaction current, and only the intercalation current contributes to updating the lithium stoichiometry in the negative electrode. Overall, Eqs. \ref{eq:dthetapdt} and \ref{eq:dthetandt} \rev{amount to a zero-dimensional, `dual-tank' representation of lithium stoichiometries at each electrode, providing a basis for tracking reaction potentials and expansions at each electrode (Figure} \ref{fig:model}B).

\subsection{Electrode Overpotentials}
\label{sec:overpotential}

The electrode surface overpotentials are governed by charge-transfer kinetics at the electrode-electrolyte interface and solid-state diffusion dynamics. Our work reduces the overpotential dynamics using a first-order representation as follows:
\begin{equation} \label{eq:etan1}
    \etai(t) = \Roi\Iapp(t) + \Rwi\Iwi(t),
\end{equation}
where $i$ is an index for the positive ($\mathrm{p}$) or negative ($\mathrm{n}$) electrode, $\etai$ is the overpotential, $\Roi$ is a lumped resistance term that includes both series and charge-transfer resistance, $\Rwi$ represents solid-state diffusion resistance, and $\Iwi$ follows first-order dynamics given by \citep{Plett2015-vo}:
\begin{equation} \label{eq:idiff}
    \frac{d\Iwi(t)}{dt} = -\frac{1}{\taui}\Iwi(t) + \frac{1}{\taui}\Iapp(t),
\end{equation}
where $\taui$ is the diffusion time constant. Note that Eqs. \ref{eq:etan1} and \ref{eq:idiff} are identical to the equations for an R-RC equivalent circuit model.

The negative electrode reaction potential, $\etan$, will play a role in determining the SEI reaction kinetics at the negative electrode-electrolyte interphase, which will be detailed in the next section.

\subsection{SEI Reaction Kinetics}
\label{sec:sei}

Since our experimental data uses ethylene carbonate (EC) as a solvent, we will describe our representation of SEI reaction kinetics in the context of EC reduction for convenience. However, it is to be understood that the formulation described here can apply to any SEI reaction that proceeds according to Eq. \ref{eq:sei-reaction}. A canonical EC reduction reaction pathway proceeds according to \citep{Aurbach1987-tl, Zhang2015-si}:
\begin{equation} \label{eq:ec}
    2\ec + 2\electron + 2\lip \rightarrow \mathrm{LEDC}\downarrow + \ethylene\uparrow,
\end{equation}
where LEDC is lithium \rev{ethylene dicarbonate}, the solid reaction product considered to be the SEI \citep{Zhuang2005-tf}, and $\ethylene$ is ethylene gas. The SEI reaction current density is assumed to take a Tafel-like form \citep{Safari2008-zk}:
\begin{equation} \label{eq:jsei-bv}
    \jsei = nF\ksei\cecs\exp\left(-\frac{\alphasei nF}{RT}\etasei\right),
\end{equation}
where $\ksei$ is the reaction rate constant, $\cecs$ is the concentration of solvent molecules at the reaction surface, $\alphasei$ is the symmetry factor, and $\etasei$ is the SEI reaction overpotential. $n$ is the number of electrons involved in the reaction. For the EC reaction, $n=2$. In the Doyle-Fuller-Newman model, $\etasei$ is determined by \citep{Fuller1994-pm, Yang2017-uf}:
\begin{equation} \label{eq:etasei}
\etasei = \phisn - \phie - \Usei - \japp\Rsei\deltasei,
\end{equation}
where $\phisn$ is the electrode surface potential, $\phie$ is the electrolyte potential, $\Usei$ is the SEI reaction potential, $\Rsei$ is the SEI resistivity, and $\deltasei$ is the SEI thickness. Note that $\phisn$ and $\phie$ do not explicitly appear in our formulation due to our simplified model representation. To then define $\etan$ in our system, we notice that \citep{Fuller1994-pm, Yang2017-uf}:
\begin{equation} \label{eq:etan}
    \etan = \phisn - \phie - \Un(\thetans) - \japp\Rsei\deltasei,
\end{equation}
which can be substituted into Eq. \ref{eq:etasei} to yield:
\begin{equation} \label{eq:etaseired}
    \etasei = \etan + \Un(\thetans) - \Usei.
\end{equation}
Eq. \ref{eq:etaseired} is now directly solvable, since $\etan$ is provided for by Eq. \ref{eq:etan1}. Note that this representation of $\etasei$ does not explicitly represent the SEI resistivity $\Rsei$. However, this is easily remedied by lumping $\Rsei$ into $\Ron$ in Eq. \ref{eq:etan1}.

Solvent transport limitations through the SEI is modeled using a linearized Fick's law which takes the form \citep{Deal1965-pb, Yang2017-uf}:
\begin{equation} \label{eq:jsei-dif}
    \Dsei\frac{\cecs-\ceco}{\deltasei} = \frac{\jsei}{nF},
\end{equation}
where $\Dsei$ is the effective diffusivity of the SEI and $\ceco$ is the concentration of solvent molecules in the bulk electrolyte phase. Reniers et al. combined Eqs. \ref{eq:jsei-bv} and \ref{eq:jsei-dif} into a more explicit representation of the reaction and diffusion-limited processes by eliminating $\cecs$ from the equations, yielding \citep{Reniers2019-rp, Sulzer2021-ud}:
\begin{equation} \label{eq:jsei-mix}
\jsei = \frac{-\ceco}{1/(nF\ksei\exp(-\alphasei nF\etasei/RT)) + \deltasei/(\Dsei nF)}.
\end{equation}
We further rewrite Eq. \ref{eq:jsei-mix} in terms \rev{of two limiting currents}:
\begin{equation} \label{eq:jsei-mix2}
    \frac{1}{\jsei} = \frac{1}{\jseirxn} + \frac{1}{\jseidif},
\end{equation}
with:
\begin{align}
    \label{eq:jsei-rxnmax}
    \jseirxn  &\triangleq nF\ksei\ceco\exp\left(-\frac{\alphasei nF}{RT}\etasei\right) \\
    \label{eq:jsei-difmax}
    \jseidif &\triangleq \frac{nF\Dsei\ceco}{\deltasei}.
\end{align}
Eq. \ref{eq:jsei-rxnmax} represents the reaction-limited SEI current in the absence of diffusion limitations ($\cecs = \ceco$) while Eq \ref{eq:jsei-difmax} represents the diffusion-limited SEI current in the absence of reaction limitations ($\cecs=0$). Eq. \ref{eq:jsei-mix2} thus highlights that the SEI current is the harmonic mean of two limiting currents. The slower of the two processes limits the overall flow of SEI current.

The SEI current density can be converted to total SEI current by:
\begin{equation} \label{eq:isei}
\Isei = \asn\An\Ln\jsei,
\end{equation}
where $\asn$ is the specific surface area (i.e. surface-to-area volume ratio) of the negative electrode, $\An$ is the geometric area of the negative electrode, and $\Ln$ is the geometric thickness of the negative electrode. The SEI current can then be directly integrated to yield the total capacity of lithium lost to SEI-forming reactions:
\begin{equation} \label{eq:qsei}
    \Qsei = \int\Isei(t)\mathrm{d}t.
\end{equation}

Current conservation can finally be used to solve for the total lithium intercalation current:
\begin{equation} \label{eq:Iint}
    \Iint = \Iapp - \Isei.
\end{equation}

Eq \ref{eq:Iint} unifies the SEI reactions during charging, discharging, and resting, in accordance with the reaction scheme proposed in Figure \ref{fig:schematic}. During charging, $\Iapp$ is positive and is split between $\Iint$ and $\Isei$ which are both positive. During discharging, $\Iapp$ and $\Iint$ are both negative, but $\Isei$ remains positive by definition since the SEI reaction is irreversible. During rest, $\Iapp = 0$ so $\Iint = -\Isei$, consistent with the scheme that lithium deintercalation from the negative electrode drives the SEI reaction during rest. In all cases, Eqs. \ref{eq:jsei-mix2} and \ref{eq:Iint} provide a consistent framework for capture the SEI reaction dynamics.

\subsection{Solvent Consumption}

As SEI grows, solvent molecules are consumed according to Eq. \ref{eq:sei-reaction}, decreasing the bulk solvent concentration. The solvent consumption process can be described by:
\begin{equation} \label{eq:solvcons}
    \frac{d\ceco}{dt} = \frac{\asn\jsei}{nF}.
\end{equation}
As solvent is consumed, $\jsei$ is further decreased according to Eq. \ref{eq:jsei-mix}. The solvent depletion process is thus self-limiting.

\subsection{Expansion Modeling}
\label{sec:expansion}

The goals of developing an expansion model are three-fold. First, an expansion model enables the prediction of macroscopic expansion trends during formation cycling and aging seen in Figure \ref{fig:experimental}, which could then be extended to predict expansion over the remaining lifetime of the cell. Second, the predicted expansion dynamics can lead to deeper insights into the formation process since SEI growth and irreversible cell expansion are inextricably linked. Finally, an expansion model fit to experimental expansion data enables a richer dataset for model parameterization.

This work proposes a phenomenological representation of the total cell thickness expansion, $\Delta_\mathrm{tot}$, of the form:
\begin{equation} \label{eq:expansion}
\Delta_\mathrm{tot}(t) = \Delta_\mathrm{SEI}(t) + \Delta_\mathrm{rev}(t).
\end{equation}
The first term in Eq. \ref{eq:expansion} represents the irreversible cell thickness expansion due to SEI film growth and is given by: 
\begin{equation} \label{eq:dlsei}
    \Delta_\mathrm{SEI}(t) = \frac{N_\mathrm{layers}\Ln}{\Rn(1+\nun(\thetan(t))/3)}\deltasei(t), 
\end{equation}
where $\deltasei$ is SEI film thickness at a negative electrode particle, $\Ln$ is the geometric length of the negative electrode, and $N_\mathrm{layers}$ is the number of active layers in the stacked configuration. $\Rn$ is the radius of a single negative electrode particle and $\nun$ is the reversible expansion function of the negative electrode (see Figure \ref{fig:ocp}). Eq. \ref{eq:dlsei} is derived in \rev{Appendix} \ref{sec:expansion-derivation}.

The SEI film thickness, $\deltasei$, evolves due to the accumulation of SEI current according to Safari et al. \citep{Safari2008-zk}:
\begin{equation} \label{eq:ddeltadt}
\frac{d\deltasei}{dt} = \vmsei \frac{\jsei}{nF}.
\end{equation}
$\vmsei$ is the SEI molar volume in m$^3$/mol, defined by  $\vmsei\triangleq\Msei/\rhosei$, where $M_\mathrm{SEI}$ is the SEI molecular weight and $\rho_{\mathrm{SEI}}$ is the SEI density.

The second term in Eq. \ref{eq:expansion} represents the reversible expansion of the positive and negative electrodes, given by:
\begin{equation} \label{eq:dlrev}
    \Delta_\mathrm{rev}(t) = N_\mathrm{layers}\cdot\left(\frac{\Lp}{3}\nup(\thetap(t)) + \frac{\Ln}{3}\nun(\thetan(t))\right),
\end{equation}
where $\nup$ and $\nun$ are the volumetric expansion functions for each electrode, shown in Figures \ref{fig:model} and \ref{fig:ocp}. The reversible volumetric expansion functions are due to lithium intercalation-induced swelling of the electrodes. Graphite, for example, expands up to 12\% volumetrically during lithiation \citep{Schweidler2018-vm}. Layered oxide materials also expand and contract, but the total expansion depends on the range of lithium stoichiometries reachable within the full cell voltage window \citep{Li2021-hv}. The expansion functions $\nup$ and $\nun$ are typically quantified by measuring unit cell lattice parameter changes during lithiation and delithiation via in-situ X-ray diffraction \citep{Schweidler2018-vm, Li2021-hv}. The prefactor terms convert the microscopic volumetric expansions to macroscopic thickness expansions and are derived in \rev{Appendix} \ref{sec:expansion-derivation}.

Further expansion modeling assumptions and clarifications are given as follows:
\begin{itemize}
    \item Expansion and compression of the inactive layers, including the separators, current collectors, and pouch cell enclosure, are ignored.
    \item Electrode particles are spherical.
    \item The expansion functions $\nup(\thetap)$ and $\nun(\thetan)$ remain invariant over the formation and aging process.
    \item Expansion and contraction in the planar direction are ignored.
\end{itemize}

\subsection{Boosted SEI Growth Rate During Charging}
\label{sec:boost}

Here, introduce a concept called ``SEI growth boosting.'' The boosting refers to enhanced SEI growth rate during cell charging which has been previously explored \rev{in the context of degradation modeling} \cite{Kupper2018-qh, Laresgoiti2015-qj, Keil2020-vo}. This effect is especially important to consider during formation cycling, during which the electrodes are experiencing the largest change in lithium stoichiometry, creating more particle-level stresses which act as a source of the enhanced SEI growth rate \cite{Keil2020-vo, Laresgoiti2015-qj}. As we will later show in Section \ref{sec:boost-explore}, this model extension was necessary for explaining the observed macroscopic formation trends from Figure \ref{fig:experimental}.

\begin{figure}[ht!]
\begin{center}
\includegraphics[width=0.7\linewidth]{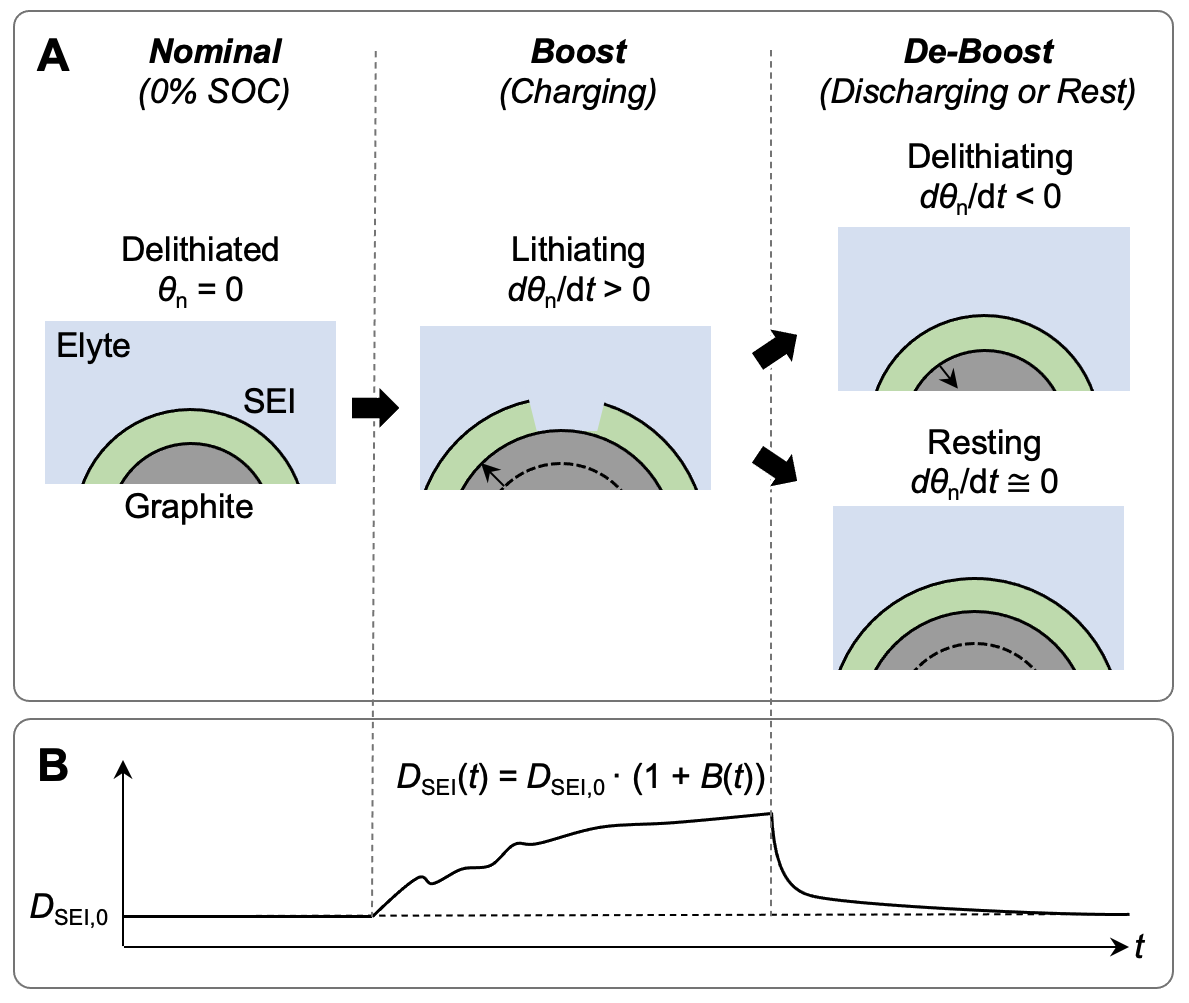}
\end{center}
\caption{\textbf{Schematic of the SEI growth boosting mechanism.} (A) During lithiation (charging), SEI growth rate is boosted due to SEI fracture as the negative electrode particles expand. During resting and delithiation (discharging), the SEI growth rate is de-boosted to the nominal rate as the SEI film `self-heals.' (B) Boosting dynamics is represented by $B(t)$ which modifies the effective SEI diffusivity $\Dsei$. Elyte: electrolyte.}
\label{fig:boost}
\end{figure}

Figure \ref{fig:boost} \rev{illustrates the boosted SEI growth mechanism}. Panel A describes a \rev{graphite} particle that starts at some initial state of lithiation $\thetan = \theta_{\mathrm{n},0}$. As the full cell is charged, the graphite lithiates and expands. \rev{Ideally,} the SEI elastically deforms to accommodate the particle swelling. However, if the SEI film is brittle, then parts of the SEI film may fracture, exposing fresh electrode surfaces to new electrolyte \citep{Laresgoiti2015-qj}, shown in Figure \ref{fig:boost}A. \rev{Reacting molecules near these newly-exposed electrode surfaces will see more facile reaction kinetics since no pre-existing SEI film is present to limit the diffusion of reacting molecules to the reaction surface. The overall SEI current density will thus be temporarily boosted.}

\rev{Next, we consider the case of delithiation and resting. During resting, new SEI is formed to fill in the fresh electrode surfaces. As the SEI thickness in these regions approach the volume-averaged thickness, the overall SEI reaction rate begins to resemble the rate prior to boosting}. During delithiation, the SEI is assumed to remain in contact with the graphite particles which are contracting, so no new electrode surfaces are exposed. Note that this assumption may be violated by systems having high volumetric expansions such as silicon \citep{Pereira2022-tq} which we leave for future work to explore. Overall, \rev{during resting and discharging, we assume that boosting no longer occurs and the SEI growth rate is gradually restored to the original rate prior to boosting.}

\rev{This work interprets} the SEI growth boosting process as a modification to the effective SEI diffusivity, $\Dsei$. In this interpretation, \rev{newly exposed electrode surfaces are represented as increases to the local SEI porosity which in turn increase the volume-averaged SEI porosity} $\varepsilon$. Changes in the volume-averaged porosity are encapsulated by the effective diffusivity according to \citep{Petersen1958-po, Chung2013-pm}:
\begin{equation} \label{eq:dseieff}
    \Dsei(t) = D_{\mathrm{SEI},0}\frac{\varepsilon(t)}{\tau},
\end{equation}
where $\Dsei$ is the effective diffusivity that we have been using for this work, $D_{\mathrm{SEI},0}$ is a reference diffusivity, $\varepsilon$ is the volume-averaged SEI film porosity, and $\tau$ is the tortuosity. Note that this expression can be further simplified using the Bruggeman relation $\tau = \varepsilon^{-0.5}$ \citep{Fuller1994-pm}.

To describe the dynamics of the porosity evolution, we define an empirical variable $B(t)$ which modifies the effective SEI diffusivity according to:
\begin{equation} \label{eq:dseiboosted}
    D_{\mathrm{SEI,boosted}}(t) = D_{\mathrm{SEI},0} (1 + B(t)),
\end{equation}
where $D_{\mathrm{SEI,boosted}}$ is the boosted SEI diffusivity and $B(t)$ can be interpreted as a boost factor. \rev{We assume that the dynamics of boosting is described by some unknown function $f(B(t))$ which is driven by the negative electrode expansion rate:}
\begin{equation}
    f(B(t)) = \gamma\frac{d\nun(\thetan)}{dt},
\end{equation}
where $\nun$ is the negative electrode volume expansion function and $\gamma$ is an input sensitivity parameter. A first-order Taylor expansion of $f(B(t))$ leads to our proposed state equation describing SEI growth boosting:
\begin{equation} \label{eq:boost}
    \tau\frac{dB}{dt} + B = \gamma\frac{d\nun}{dt}.
\end{equation}
In this equation, $\tau$ is the time constant for the first-order dynamics. This equation can be further separated for boosting during lithiation and ``de-boosting'' during delithiation and rest, with separate time constants describing each process:
\begin{equation} \label{eq:boost2}
\left\{
    \begin{array}{rlll}
      \tau_{\uparrow}\frac{dB}{dt} + B &= \gamma\frac{d\nun}{dt} & \Iapp > 0 & \text{(Boost)}\\
      \tau_{\downarrow}\frac{dB}{dt} + B &= 0 & \Iapp \leq 0 & \text{(De-boost)}.
    \end{array}
\right.
\end{equation}
During charging, the boosting time constant $\tau_\uparrow$ describes how quickly the effective diffusivity increases in response to newly-created reaction surfaces for SEI growth. During discharging and resting, the de-boosting time constant $\tau_\downarrow$ describes the rate of ``self-healing'' as the freshly-created surfaces fill up with new SEI and the effective diffusivity approaches its original value, $D_\mathrm{SEI,0}$. 

\subsection{Homogenized Multi-Species SEI Reaction Model}
\label{sec:homogenized}

\begin{figure}[ht!]
\begin{center}
\includegraphics[width=1.0\linewidth]{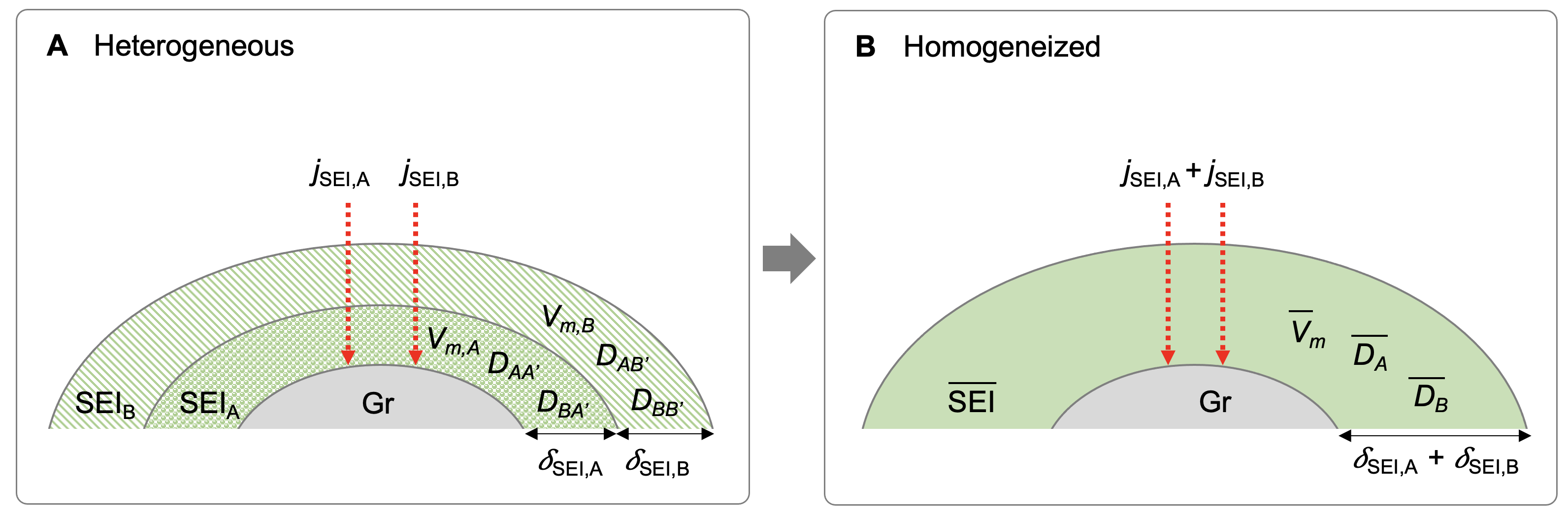}
\end{center}
\caption{\textbf{Homogenized representation of multi-species SEI reaction kinetics.} (A) Heterogeneous representation. (B) Homogenized representation. Variables with bars indicate volume-averaged (homogenized) quantities. In the homogenized representation, SEI reactions occur in parallel but the reaction kinetics are coupled through a homogenized medium represented by homogenized diffusivities for each reacting species, $r$, $\Dseirbar$, and the total SEI thickness. In the heterogeneous case, there are ($n_r \times n_r$) diffusivities, where $n_r$ is the number of reacting species. In the homogeneous case, the number of diffusivities is reduced to only $n_r$ elements.}
\label{fig:homogenized}
\end{figure}

\rev{So far, our description of SEI growth reaction kinetics  applied to the case of a single reacting species forming a single-component SEI solid product. However, the SEI growth process in commercially-relevant systems involves the simultaneous reaction of multiple electrolyte components including solvent and additive components. We} therefore extend our model to describe the case of multiple reacting species, using Figure \ref{fig:homogenized} as a guide. This model extension enables a reduced-order representation of multiple, parallel SEI reactions. We will later demonstrate the usage of this model to represent the reaction of two electrolyte components, a solvent species (EC) as well as an additive species (VC), during the first formation charge cycle (see Section \ref{sec:first-charge}). 

This model extension treats each SEI reaction as being governed by its own set of rate parameters (e.g. $\Usei, \Dsei$, $\ksei$, ...). However, the reactions kinetics become coupled since all reacting species must diffuse through the same set of solid SEI products to reach the reaction interface. Moreover, at the reaction interface, the charge-transfer reaction kinetics also become coupled since the chemical state of the electrode is described by $\Un$, which is a singular value which evolves as a function of the total SEI reaction current. A mathematical treatment is as follows.

Each SEI reaction proceeds according to Eq. \ref{eq:sei-reaction} and is assumed to occur in parallel. The total SEI reaction current density thus takes the form:
\begin{align} \label{eq:jsei-multi}
    \jsei &= \sum_r{\jseir} \\
          &= \sum_r{\left(\frac{1}{\jseirxnr} + \frac{1}{\jseidifr}\right)^{-1}},
\end{align}
where $r = \{\mathrm{EC}, \mathrm{VC}, ...\}$ represents the different electrolyte species that are reduced to form their respective SEI solid products, and the expansion of the right-hand side is due to Eq. \ref{eq:jsei-mix2}. 

The reaction-limited \rev{current density} for the $r$th species is:
\begin{equation} \label{eq:jseirxnr}
    \jseirxnr = n_r F\kseir\cseio\exp\left(-\frac{\alphaseir n_r F}{RT}\left(\etan + \Un(\thetans) - \Useir\right)\right).
\end{equation}
Each reaction is thus governed by independent rate parameters $\{\kseir, \cseir, \Useir, n_r\}$. \rev{However,} a coupling is introduced through $\Un(\thetans)$ which changes as a function of the total SEI current.

To model the diffusion-limited current densities, the diffusion parameters in Eq. (\ref{eq:jsei-dif}) require reinterpretation, since each reacting species $r$ must now diffuse through a solid SEI layer consisting of multiple solid reaction products. Here, we update our interpretation of the effective diffusivity to consider both the reacting molecule and its environment:
\begin{equation}
    D_{\mathrm{SEI},rl}: \hbox{reacting molecule $r$ through solid product $l$}.
\end{equation}
With $n_r$ reacting molecules and $n_l$ resulting solid products, we assume that $n_r = n_l$ according to Eq. \ref{eq:sei-reaction}, and therefore, $D_{\mathrm{SEI},rl}$ is a matrix of size ($n_r \times n_r$). To simplify the model, we introduce the notion of average diffusivities for each reacting species, according to:
\begin{equation} \label{eq:deff}
    \frac{1}{\Dseirbar} = \sum_l\frac{\mu_l}{\Dseirl},
\end{equation}
where $\Dseirbar$ is the average diffusivity of reacting species $r$ through a homogenized solid medium consisting of $l$ solid products. $\mu_l = m_l/\sum_l{m_l}$ are weights based on the mass of each solid product $m_l$. In this representation, each reacting species $r$ is assigned its own diffusivity which \rev{describes} the diffusivity of species $r$ through a homogenized environment consisting of $n_l$ solid products. The model thus carries only $n_r$ number of $\Dseirbar$.

The diffusion-limited SEI current density can then be written as:
\begin{equation} \label{eq:jseidifr}
    \jseidifr = \frac{\Dseirbar\cseio n_r F}{\deltaseibar}.
\end{equation}

The thickness of each SEI layer grows independently according to:
\begin{align} \label{eq:growth}
    \frac{d\deltaseir}{dt} &= \vmseir\frac{\jseir}{n_r F}.
\end{align}
The total SEI thickness is taken to be the sum of each individual thickness:
\begin{equation} \label{eq:deltaseibar}
    \deltaseibar = \sum_r\deltaseir.
\end{equation}
In this representation, the thickness of each SEI is defined in a volume-averaged sense without making a distinction on the spatial arrangement of the SEI layers. The model thus remains zero-dimensional.

\rev{In summary, the multi-species SEI reaction model treats the solid SEI product as a zero-dimensional, homogenized medium with volume-averaged properties. While each SEI reaction is described by independent sets of rate parameters, the reactions become coupled since each reacting molecule must diffuse through the same medium and react at the same electrode surface. The homogenization approach taken for this work enables the prediction of non-trivial SEI formation dynamics involving multiple electrolyte components, as will be detailed in Section} \ref{sec:results}.

\section{Model Implementation}
\label{sec:parameters}

Equations from Table \ref{tbl:equations} were numerically integrated using a forward difference scheme. The simulation was run with a timestep of $\Delta t$ = 10 seconds. Each charge and discharge cycle takes less than 1 second to complete execution on a 2.6 GHz 6-Core Intel Core i7 processor. All code was written in Python and is publicly available at \url{github.com/wengandrew/formation-modeling}. The code is written to be modular and extensible, drawing from the design of other battery simulation software suites \citep{Sulzer2021-eg, Reniers2023-bg}.

The simulation framework also supports constant voltage (i.e. potentiostatic) operation mode. During this charging mode, Eq. \ref{eq:vt} is inverted to update the current for a given target voltage. The reduced-order representation of cell overpotentials allows for a closed-form solution of the form:

\begin{equation} \label{eq:icv}
I_\mathrm{CV} = \frac{\Vt - \Up + \Un - \sum_i \Rwi\Iwi\exp{(-\frac{\Delta t}{\taui})}}{\sum_i\left(\Roi + \Rwi(1-\exp{(-\frac{\Delta t}{\taui})})\right)},
\end{equation}
where $\taui\triangleq\Rwi\Cwi$.

\begin{table}[htb]
\caption{Equations describing the formation model.}
\label{tbl:equations}
\vspace*{+0.5ex}
\begin{spreadlines}{3pt}
\setlength{\belowdisplayskip}{-0pt}
\begin{flalign}
\notag\\[-2\jot]
\toprule\addlinespace
& \textsf{Electrode dynamics} & \frac{d\thetap(t)}{dt} &= -\frac{\Iapp(t)}{\Qp} \qquad\qquad
                                \frac{d\thetan(t)}{dt} = +\frac{\Iint(t)}{\Qn} & (\ref{eq:dthetapdt},\ref{eq:dthetandt})\nonumber \\
& & \Vt &= \Up(\thetaps) - \Un(\thetans) + \etap + \etan & (\ref{eq:vt}) \nonumber \\ 
& & \etai(t) &= \Roi\Iapp(t) + \Rwi\Iwi(t) \qquad\qquad i\in(\mathrm{n},\mathrm{p}) & (\ref{eq:etan1}) \nonumber\\ 
& & \frac{d\Iwi(t)}{dt} &= -\frac{1}{\taui}\Iwi(t) + \frac{1}{\taui}\Iapp(t) & (\ref{eq:idiff}) \nonumber\\ 
& \textsf{SEI reactions (EC only)} & \Iint &= \Iapp - \asn\An\Ln\jsei & (\ref{eq:isei}, \ref{eq:Iint}) \nonumber\\
& & \jsei &= \left(\frac{1}{\jseirxn} + \frac{1}{\jseidif}\right)^{-1} & (\ref{eq:jsei-mix2}) \nonumber\\ 
& & \jseirxn  &= nF\ksei\ceco\exp\left(-\frac{\alphasei nF}{RT}\left(\etan + \Un(\thetan) - \Usei\right)\right) & (\ref{eq:jsei-rxnmax}) \nonumber\\
& & \jseidif &= \frac{nF\Dsei\ceco}{\deltasei} & (\ref{eq:jsei-difmax}) \nonumber\\ 
& \textsf{Solvent consumption} & \frac{d\ceco}{dt} &= -\frac{\asn\jsei}{nF} & (\ref{eq:solvcons}) \nonumber\\ 
& \textsf{Expansion} & \Delta_\mathrm{tot} &= c_0\deltasei + c_1\nup(\thetap) + c_2\nun(\thetan) \qquad \text{(simplified)} & (\ref{eq:expansion-simplified}) \nonumber\\
& & \frac{d\deltasei}{dt} &= \vmsei\frac{\jsei}{nF} &(\ref{eq:ddeltadt}) \nonumber\\
& \textsf{SEI boosting} & D_{\mathrm{SEI,boosted}} &= D_{\mathrm{SEI,0}} (1 + B) & (\ref{eq:dseiboosted}) \nonumber\\
& & \tau_{\uparrow}\frac{dB}{dt} + B &= \gamma\frac{d\nun}{dt} \qquad \text{ Charging (boost)} & (\ref{eq:boost2}) \nonumber\\ 
& & \tau_{\downarrow}\frac{dB}{dt} + B &= 0 \qquad\qquad \text{Discharging and rest (de-boost)} & (\ref{eq:boost2}) \nonumber\\ 
& \textsf{Multi-species SEI} & \jsei &= \sum_r \jseir = \sum_r{\left(\frac{1}{\jseirxnr}+\frac{1}{\jseidifr}\right)^{-1}} & (\ref{eq:jsei-multi}) \nonumber\\
& & \jseirxnr &= n_r F\kseir\cseio\exp\left(-\frac{\alphaseir n_r F}{RT}\left(\etan + \Un(\thetans) - \Useir\right)\right) & (\ref{eq:jseirxnr}) \nonumber\\
& & \jseidifr &= \frac{\Dseirbar\cseio n_r F}{\deltaseibar} & (\ref{eq:jseidifr}) \nonumber\\
& & \Dseirbar &= \left(\sum_l\frac{\mu_l}{\Dseirl}\right)^{-1} &(\ref{eq:deff}) \nonumber\\ 
& & \deltaseibar &= \sum_r\deltaseir & (\ref{eq:deltaseibar}) \nonumber\\
\addlinespace\bottomrule
\notag
\end{flalign}
\end{spreadlines}
\end{table}

\textit{Parameterization.} Model parameters used for this work are summarized in Table \ref{tab:param}. Model parameters were either calculated based on the cell build parameters, taken from literature, fit to the experimental data, or assumed. \rev{Since parameter identification was not a focus of this work, the parameters chosen for this work may not be optimal. A more detailed investigation of parameter identification and output sensitivity analysis will be left for future work.}

\textit{SEI components.} While the formation model supports simulating an arbitrary number of SEI-reacting species, we focus on studying a model system consisting of two SEI components: ethylene carbonate (EC) and vinylene carbonate (VC). These two components represent a subset of the experimentally-tested electrolyte (3:7 EC:DEC with 2 wt\% VC, see Section \ref{sec:cellbuild}). Both EC and VC are also well-studied in literature, with established reaction pathways and some known model parameters \cite{Jeong2001-bv, Huang2019-qh, Attia2019-jg, Attia2021-bl}. VC, in particular, is present in a small quantity (2 wt\%) as an electrolyte additive and is known to be consumed during the formation process to help stabilize the SEI film \cite{Pritzl2017-wa, Aurbach2002-zq, Haruna2016-nn, Wang2002-aq}. VC, which reduces at ca. 1.35V \citep{Jeong2001-bv}, is also markedly higher than EC, which reduces at ca. 0.8V \citep{Huang2019-qh}, and is thus expected to reduce first. VC thus provides a contrasting solvent component to understand both the capabilities and limitations of the model's ability to capture first-cycle reaction dynamics with multiple SEI components. EC and VC starting concentrations were calculated to match the experimental values. The EC reaction is given by Eq. \ref{eq:ec}. The VC reaction is given by \citep{Ota2004-nl}:
\begin{equation} \label{eq:vc}
    2\mathrm{VC} + 2\electron + 2\lip \rightarrow \mathrm{LVDC}\downarrow + \ethylene\uparrow,
\end{equation}
where LVDC is lithium vinylene dicarbonate, the VC-derived SEI layer. Like EC and other solvents, multiple reaction pathways are possible, and each reaction may occur in multiple stages \citep{Ota2004-nl}. However, for this work, we simplify the representation by considering only Eq. \ref{eq:vc} as the VC reaction scheme.

\textit{Electrode capacities.} Electrode capacities $\Qp$ and $\Qn$ were initialized using a differential voltage fitting procedure outlined in Weng et al. \cite{Weng2023-lj}. This method uses a slow-rate (i.e. C/20) voltage curve to extract information about electrode-specific capacities. The C/20 charge curve used for this analysis was taken at the end of the two-week formation aging step. \rev{This work assumes that the electrode capacities remain invariant throughout the formation process, i.e. no loss of active material. This assumption can be revisited as part of future work.}

\textit{Initial conditions.} Special attention was made to set the initial cell state prior to formation cycling. Before the formation cycles, we assumed that the negative electrode was fully delithiated ($\thetan=0$) and the positive electrode was fully lithiated ($\thetap=1$). The measured cell terminal voltage before formation was then used to constrain the set of possible half-cell potentials, according to:
\begin{equation} \label{eq:voltage-bc}
    V_t|_{t=0} = \Up(\thetan=1)|_{t=0} - \Un(\thetap=0)|_{t=0}.
\end{equation}
For the cell under study, the full cell terminal voltage $V_t$ was measured to be 635mV before formation charging began. To identify the initial potentials of the positive and negative electrodes before formation, half-cells with lithium metal counter-electrodes were built. Half-cell potentials were measured immediately after electrolyte fill and assembly. The negative electrode (graphite) half-cells were measured to have potentials of $\{2.939, 2.925, 0.770, 1.698\}$V vs Li/Li$^+$, and the positive electrode (NMC622) half-cells were measured to have potentials of $\{1.814, 2.813\}$V vs Li/Li$^+$. The wide range of measured potentials measured suggest that more work is necessary to verify the initial half-cell potentials which we leave as future work. For this work, we picked $\Un(\thetan=0)=2.00V$ and $\Up(\thetap=1)=2.65V$, which satisfies Eq. \ref{eq:voltage-bc}.

\textit{Half-cell potentials.} Half-cell near-equilibrium potential functions $\Un$ and $\Up$ were adapted from Mohtat et al. \cite{Mohtat2020-zp}. These curves were linearly extrapolated to satisfy the boundary condition given in Eq. \ref{eq:voltage-bc}. These functions are plotted in Figure \ref{fig:ocp}.

\textit{Expansion functions.} Volumetric expansion functions for the positive and negative electrodes were taken from Mohtat et al. \citep{Mohtat2020-zp} and shown in Figure \ref{fig:ocp}. For the simulations, a simplified version of Eq. \ref{eq:expansion} was implemented \rev{assuming constant and fitted prefactors, i.e.:}
\begin{equation}
    \label{eq:expansion-simplified}
    \Delta_\mathrm{tot} = c_0\deltasei + c_1\nup(\thetap) + c_2\nun(\thetan).
\end{equation}
A future iteration of this work will implement Eq. \ref{eq:expansion} directly by tuning $\Ln$, $\Lp$, and $\Rn$, creating additional constraints used for model parameter tuning.

\begin{table*}[ht!]
    \caption{\textbf{Variables, parameters, and initial conditions.} Asterisks (*) indicate values computed during simulation run-time. ``I.C.'' indicates an initial condition. Double asteriks (**) indicate that values were modified from the literature values.}
    \begin{center}
    \label{tab:param}
    \begin{tabular}{l l c c l c}
        \toprule
        Symbol & Definition & Value 1 &  Value 2 & Units & Ref. \\
        \midrule
        &\textbf{SEI Properties}&$A$&$B$&& \\
        $r$ & SEI reacting species & EC & VC & - & - \\
        $l$ & SEI solid product from species $r$ & $\liedc$ & LVDC & - & - \\
        $\Useir$ & SEI reaction potential & 0.8 & 1.35 & V & Refs. \cite{Huang2019-qh, Jeong2001-bv} \\
        $\jseir$ & SEI reaction current density & * & * & A/m$^2$ & Eq. \ref{eq:jsei-mix} \\
        $\cseio$ & Bulk-phase concentration & 4541 & 304.4 & mol/m$^3$ & Ref. \citep{Safari2008-zk}\\
        $\kseir$ & Reaction rate constant & $3.0\times10^{-17}$ & $7.0\times10^{-19}$ & m/s & Fitted \\
        $\DseirA$ & Effective diffusivity of $r$ in solid $l=A'$ & $4.2\times10^{-20}$ & $4.2\times10^{-20}$ & m$^2$/s & Fitted \\
        $\DseirB$ & Effective diffusivity of $r$ in solid $l=B'$ & $6.6\times10^{-18}$ & $6.6\times10^{-18}$ & m$^2$/s & Fitted \\
        $\Qseir$ & SEI capacity of the $r$th species & * & * & A s & Eq. \ref{eq:qsei}\\
        $\vmseir$ & SEI molar volume & $9.585\times10^{-5}$ & $5.810\times10^{-5}$ & m$^{3}$/mol & Ref. \citep{Safari2008-zk} \\
        $\tau_\uparrow$ & SEI growth boosting time constant & 10 & 10 & mins & Fitted \\
        $\tau_\downarrow$ & SEI growth de-boosting time constant & 100 & 100 & mins & Fitted \\
        $\gamma$ & SEI boosting input sensitivity factor & $2.4\times10^7$ & $2.4\times10^7$ & s/m & Fitted \\
        \midrule
        &\textbf{Electrode Potentials}&\textit{pos.}&\textit{neg.}&& \\
        $\Ui$ & electrode potential vs Li/Li$^+$ & 2.635 (I.C.) & 2.000 (I.C.) & V & Fitted \\
        $\thetai$ & electrode lithium stoichiometry & 1.0 (I.C.) & 0.0 (I.C.) & - & Ref. \citep{Mohtat2020-zp}** \\
        $\Qi$ & electrode maximum capacity & 2.95 & 3.14 & Ah & Based on \cite{Weng2023-lj}\\
        \midrule
        &\textbf{Electrode Overpotentials}&\textit{pos. }&\textit{neg.}&& \\
        $\etai$ & electrode over-potential vs Li/Li$^+$ & * & * & V & Eq. \ref{eq:etan1} \\
        $\Iwi$ & electrode diffusion current (through $\Roi$) & * & * & A & Eq. \ref{eq:idiff} \\
        $\Roi$ & electrode charge-transfer resistance & 0.01 & 0.01 & $\Omega$ & Assumed \\
        $\Rwi$ & electrode diffusion resistance & 0.001 & 0.001 & $\Omega$ & Assumed \\
        $\Cwi$ & electrode diffusion capacitance & $7.6\times10^4$ & $7.6\times10^4$ & F & Assumed \\
        $\taui$ & electrode diffusion time constant & * & * & s & Assumed \\
        \midrule
        &\textbf{Electrode Geometry and Expansion}&\textit{pos.}&\textit{neg.}&& \\
        $a_{\mathrm{s},i}$ & electrode specific surface area ($3\varepsilon_i/R_i$) & - & $1.05\times10^5$ & 1/m & Assumed \\
        $A_i$ & electrode geometric area & - & 0.097566 & m$^2$ & Calculated \\
        $L_i$ & electrode geometric length & - & 80 & $\mu$m & Assumed \\
        $c_0$ & expansion fitting coefficient & - & 127 & - & Fitted \\
        $c_1$ & expansion fitting coefficient & - & 0.00045 & - & Fitted \\
        $c_2$ & expansion fitting coefficient & - & 0.00045 & - & Fitted \\
        $\Delta_\mathrm{tot}$ & macroscopic pouch cell thickness expansion & - & - & m & Eq. \ref{eq:expansion} \\
        $\nu_i$ & electrode particle volumetric expansion & * & * & \% & Ref. \citep{Mohtat2020-zp} \\
        $\deltasei$ & SEI thickness expansion & - & $5\times10^{-9}$ (I.C.) & m & Eq. \ref{eq:ddeltadt}  \\
        \bottomrule
    \end{tabular}
    \end{center}
\end{table*}

\section{Model Results}
\label{sec:results}

\rev{The formation model was simulated by inputting the current profile from Figure} \ref{fig:experimental}. The results highlight some of the key characteristics of a two-component SEI growth mechanism involving the sequential reduction of VC followed by EC. The observed results underscore the coupled nature of SEI formation dynamics, including the interplay between cell expansion and boosted SEI growth, as well as the interplay between reaction dynamics involving multiple SEI components. The same results introduced here will also set the stage for comparison against experimental data which will be discussed in Section \ref{sec:experimental}.

\subsection{SEI Growth Dynamics During First Charge}
\label{sec:first-charge}

\begin{figure}[ht!]
\begin{center}
\includegraphics[width=0.7\linewidth]{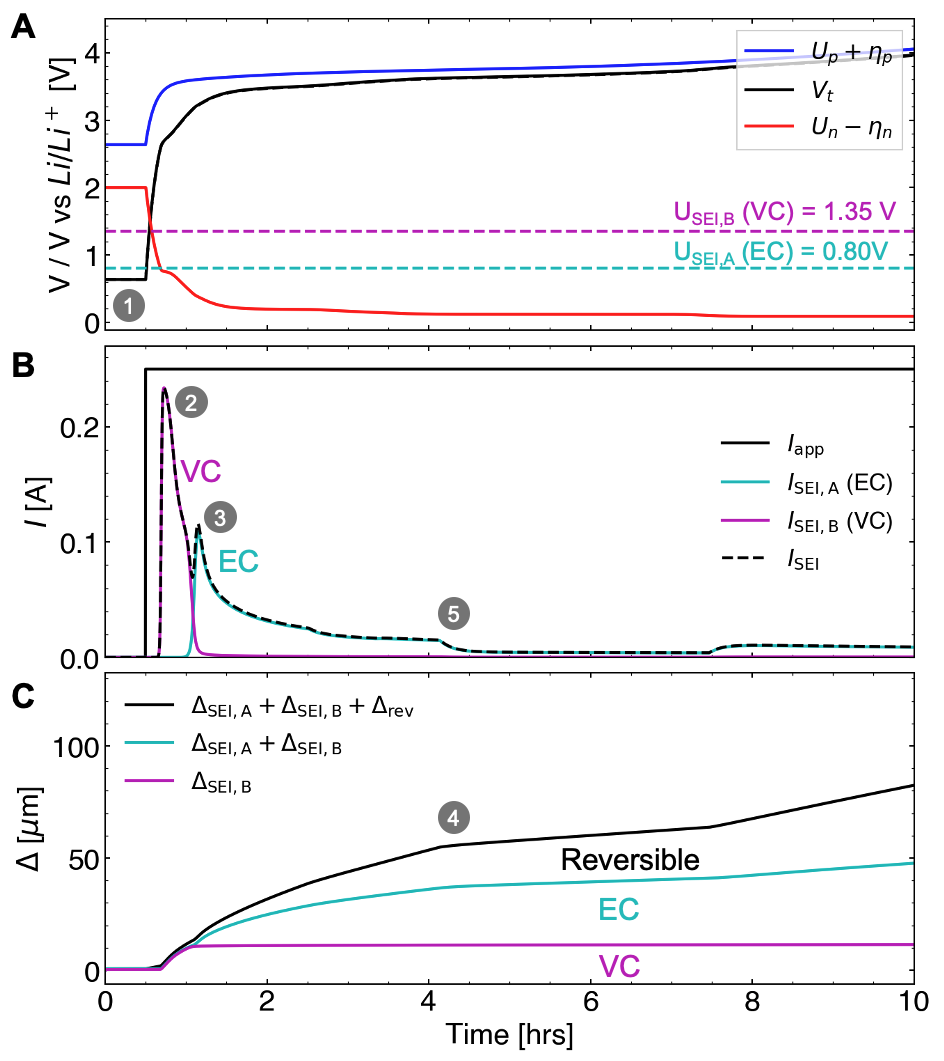}
\end{center}
\caption{\textbf{SEI reaction and full cell expansion dynamics during the first formation charge cycle.} (A) Full cell voltage, electrode potentials, and SEI reaction potentials. (B) Applied current and SEI reaction currents. (C) Full cell thickness expansion, including both reversible expansion from electrode intercalation and irreversible expansion from SEI components. EC: ethylene carbonate. VC: vinylene carbonate.}
\label{fig:model-first-charge}
\end{figure}

Figure \ref{fig:model-first-charge} highlights key model outputs before and during the first formation charge. The simulation begins with the cell at rest before any external charge is applied. At this stage, the positive electrode is completely lithiated ($\thetap=1$) and the negative electrode is completely delithiated ($\thetan=0$). Negligible SEI growth occurs during this stage since the negative electrode potential of 2.0V vs Li/Li$^+$ exceeds the reaction potentials of both VC and EC (Panel A, marker 1). After 30 minutes, an external current of C/10 (0.25A) is applied to begin the formation charge process (Panel B). Within minutes, the negative electrode potential drops below the VC reaction potential of 1.35V and VC begins to reduce (marker 2). This reaction current reaches a peak within minutes and falls back down. Next, as the negative electrode becomes more lithiated, the negative electrode potential falls below the EC reaction potential of 0.80V and EC begins reducing (marker 3). Similar to VC, the reaction current for VC also soon reaches a peak, but then gradually decreases. 

The buildup of VC and EC-derived SEI products during the initial stages of formation contributes to the macroscopically-observed cell thickness expansion (Panel C). The predicted total expansion is a result of both the irreversible expansion due to SEI growth and the reversible expansion due to lithium intercalation/deintercalation in the electrodes. The model predicts that SEI growth plays a dominant role in determining the total cell expansion during the first cycle.

During the first charge cycle, the negative electrode expands, activating the SEI growth boosting mechanism which increases the effective SEI diffusivity (see Section \ref{sec:boost}). Since the negative electrode expansion increases monotonically during the charge cycle, boosting persists throughout the cycle. Note that, at mid-SOCs, the expansion rate for the lithiated graphite is lowered temporarily (marker 4) which decreases the boost magnitude according to Eq. \ref{eq:boost2}. The EC reduction current decreases accordingly (marker 5). The boosting mechanism will be explored in more detail in Section \ref{sec:boost-explore}.

\subsection{Exploring the Limiting Current}

\begin{figure}[ht!]
\begin{center}
\includegraphics[width=0.7\linewidth]{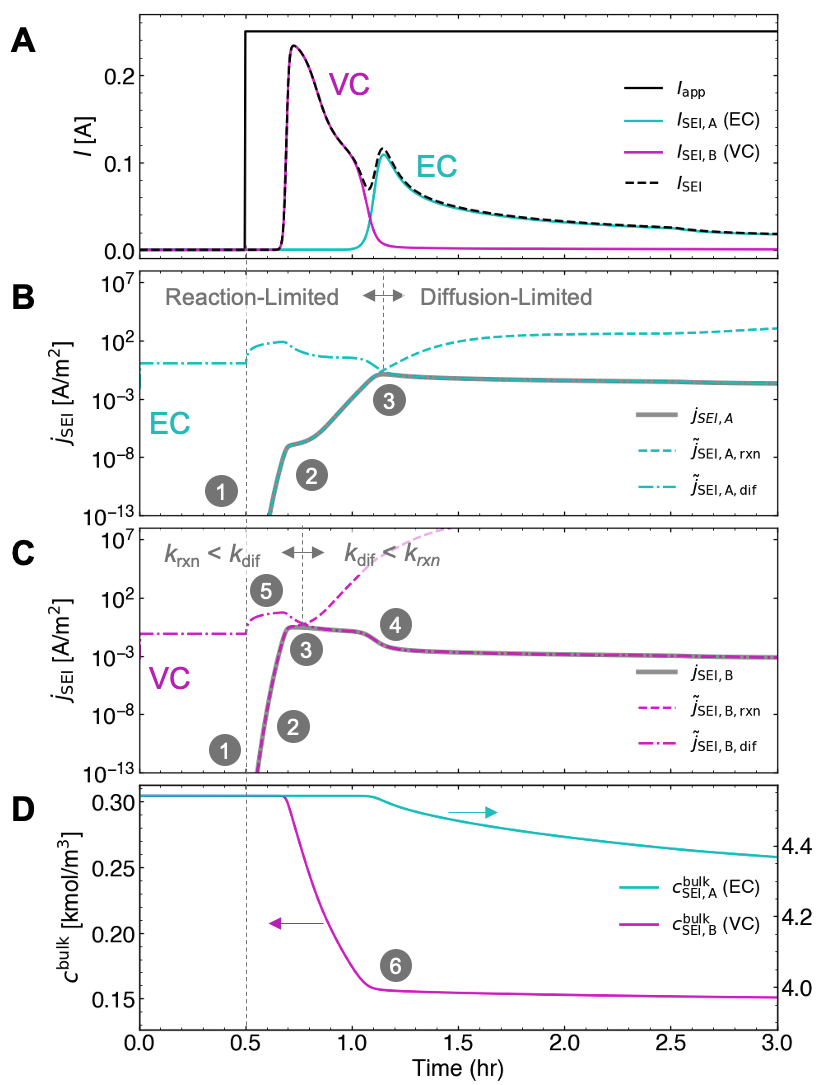}
\end{center}
\caption{\textbf{Reaction-diffusion and solvent consumption dynamics during the first formation charge cycle.} (A) Applied current and SEI reaction currents. (B) Breakdown of limiting current for EC. (C) Breakdown of limiting current for VC. (D) Bulk solvent concentration.}
\label{fig:model-diffusion-reaction}
\end{figure}

Figure \ref{fig:model-diffusion-reaction} explores the limiting SEI current during the initial stages of formation. Before any external current is applied, the SEI currents are reaction-limited since the negative electrode, at 2.0V vs Li/Li$^+$, is above the reaction potential for both VC (1.35V) and EC (0.8V) (marker 1 in both Figures \ref{fig:model-first-charge} and \ref{fig:model-diffusion-reaction}), so Eq. \ref{eq:jsei-bv} would predict very small current densities. However, after the external current is applied, the negative electrode potential decreases rapidly due to the steepness of the negative electrode equilibrium potential function $\Un$. This process pushes the SEI reaction overpotential (Eq. \ref{eq:etaseired}) towards more negative values, increasing the overall SEI current (Eq. \ref{eq:jsei-bv}). This process explains why the initial surge in SEI current coincides with the rapid decrease in the negative electrode potential (marker 2 in both Figures \ref{fig:model-first-charge} and \ref{fig:model-diffusion-reaction}). The reacting species with the higher reaction potential will tend to react first, as was the case with VC and EC in this model system.

\subsubsection{Transition from Reaction-Limited to Diffusion-Limited SEI Growth}
\label{sec:reaction-diffusion}

Shortly after the initial surge in SEI current, the SEI currents reach a peak and then begin to decrease. This can be observed for both the VC and EC reactions (marker 3 in Figure \ref{fig:model-diffusion-reaction}). We attribute this SEI current decay to the transition from a reaction-limited regime to a diffusion-limited regime. During this transition, the reaction-limited current continues to increase as the negative electrode potential continues to decrease. However, simultaneously, the diffusion-limited current decreases as the SEI film thickness increases according to Eq. \ref{eq:ddeltadt}. This film thickness increase limits the concentration of reacting molecules at the electrode surface. Eventually, the diffusion-limited current drops below the reaction-limited current. After this point, further increases in the reaction kinetics are no longer fully realized; the total reaction current becomes determined by the diffusion-limited current (Eq. \ref{eq:jsei-difmax}). The diffusion-limited current continues to decrease as the SEI film builds, limiting the overall SEI reaction rate.

\subsubsection{Multi-Species Diffusive Coupling and Boosted SEI Growth} 

After 1.2 hours, the VC reaction current density experiences another drop (marker 4 in Figure \ref{fig:model-diffusion-reaction}). This drop is due to the coupling between the EC and VC diffusivities introduced by the multi-species reaction model (see Section \ref{sec:homogenized}). As EC begins to react, the EC-derived film begins to grow. For this simulation, the effective diffusivity of the EC-derived film was assumed to be two orders of magnitude lower than that of the VC-derived film (see Table \ref{tab:param}). Thus, as soon as the EC reaction begins, the effective diffusivity of VC through the SEI, which now consists of both the VC and EC-derived films, is pulled towards the lower diffusivity of the EC-derived film, causing the VC reaction rate to be further slowed.

The effect of the boosted SEI growth mechanism is again observed in Figure \ref{fig:model-diffusion-reaction}B,C. The diffusion-limited currents here do not monotonically decrease as would be predicted by Fick's law. Rather, the diffusion-limit current initially surges when external current is applied (marker 5). This surge is explained by the negative electrode expansion which modifies the SEI diffusivity according to Eqs. \ref{eq:dseiboosted} and \ref{eq:boost}. The diffusion-limited current decreases shortly thereafter as the SEI film thickness begins to increase. See Section \ref{sec:boost-explore} for a further discussion on SEI boosting.

\subsubsection{Solvent Consumption}

Figure \ref{fig:model-diffusion-reaction}D plots bulk concentrations for EC and VC. Solvent and additive consumption occur due to SEI growth according to Eq. \ref{eq:solvcons}. The initial concentration of VC is much lower than that of EC (304 mol/m$^3$ versus 4541 mol/m$^3$) since VC is an electrolyte additive. VC is consumed rapidly during the initial stages of the first charge cycle. However, VC consumption appears to plateau afterwards (marker 6). This plateau coincides with the start of EC reduction, which depresses the VC reduction rate due to the diffusive coupling as previously explained. Hence, the VC concentration does not drop completely to zero during formation charge due to diffusion limitations. The model thus predicts the presence of trace amounts of electrolyte additives after formation cycling is completed.

\subsection{Understanding SEI Growth Boosting}
\label{sec:boost-explore}

\begin{figure}[ht!]
\begin{center}
\includegraphics[width=0.7\linewidth]{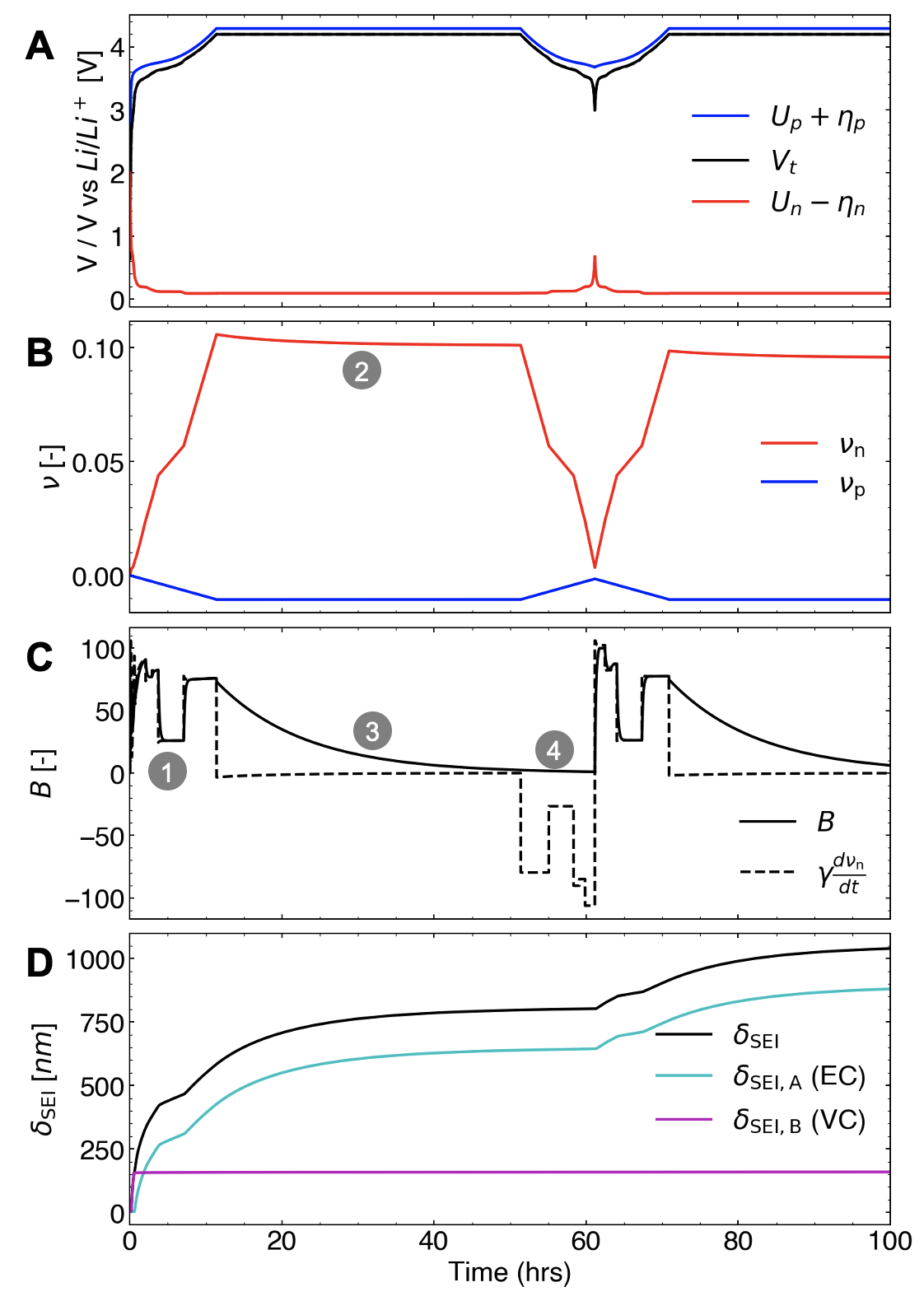}
\end{center}
\caption{\textbf{Demonstration of SEI growth rate boosting during charging and de-boosting during resting and discharging.} A simulation was configured to alternate between cycling and calendar aging, illustrating the boosted SEI growth rate during cycling charge events. (A) Full cell voltage and electrode potentials. (B) Positive and negative electrode volumetric expansion functions. (C) Boost function $B(t)$ plotted against the boosting input term $\gamma(d\nun/dt)$. (D) Comparison of SEI thickness growth rates on cycling versus calendar aging.}
\label{fig:model-boost}
\end{figure}

Figure \ref{fig:model-boost} highlights the impact of boosting on the SEI growth rate. For this simulation, the cell was first charged at C/10 to 100\% SOC, rested for 40 hours, then discharged at C/10 (Panel A). This pattern was then repeated. During each charge and discharge cycle, the negative electrode expands and contracts (Panel B). The negative electrode volumetric expansion rate, $d\nun/dt$, provides a non-linear input which activates the boosted SEI mechanism (see Section \ref{sec:boost}).

Panel C plots $B(t)$ along with $\gamma (d\nun/dt)$. During charging, $B(t)$ and $\gamma(d\nun/dt)$ track closely to each other since we have assumed a small value for the boosting time constant, with $\tau_\uparrow = 10$ mins. The step transitions (marker 1) are due to the piecewise-linear construction of the negative electrode expansion function, $\nun$, consisting of three regions: steep, shallow, and steep, corresponding to low, medium, and high lithium stoichiometries, respectively (see Figure \ref{fig:ocp}). The SEI thickness growth rate tracks these different regions accordingly, initially starting fast, then slowing down, then becoming fast again (Panel D).

During rest, $\gamma(d\nun/dt)$ approaches zero, but it is not exactly zero, since lithium continues to deintercalate from the negative electrode to form SEI during rest, causing the negative electrode to contract slightly (marker 2). Overall, the magnitude of $d\nun/dt$ during resting remains small. $B(t)$, however, takes hours to decay from its boosted state since we assumed that the de-boosting process is much slower than the boosting process, with  $\tau_\downarrow = 100$ mins (marker 3).

As the rest step transitions into a discharge step, $\gamma(d\nun/dt)$ inverts sign, but $B(t)$ continues to exponentially decay towards zero (marker 4). Physically, this behavior represents the assumption that, as the negative electrode particles contract, the SEI maintains physical contact with the particle and no new reaction surfaces are exposed. As the previously-exposed negative electrode surfaces rebuild SEI, the SEI effective diffusivities return to their nominal values prior to boosting.

Panel D highlights the effect of boosted SEI growth on SEI thickness expansion. During charging, the SEI build-up is accelerated. However, during resting and discharging, the SEI growth rate begins to plateau, corresponding to a steady-state diffusion regime. During the subsequent charge, this steady-state regime is interrupted as the SEI growth rate is boosted again by the charging process.

\section{Comparison to Experiment}
\label{sec:experimental}

A main goal of this work was to develop a model capable of capturing macroscopic trends observed in experimental formation data. Here, we demonstrate that our formation model achieves this goal for a range of metrics, including  full cell voltages, cell thickness expansions, coulombic efficiencies, and first cycle dQ/dV.

\subsection{Full Cell Voltage}

\begin{figure}[ht!]
\begin{center}
\includegraphics[width=\linewidth]{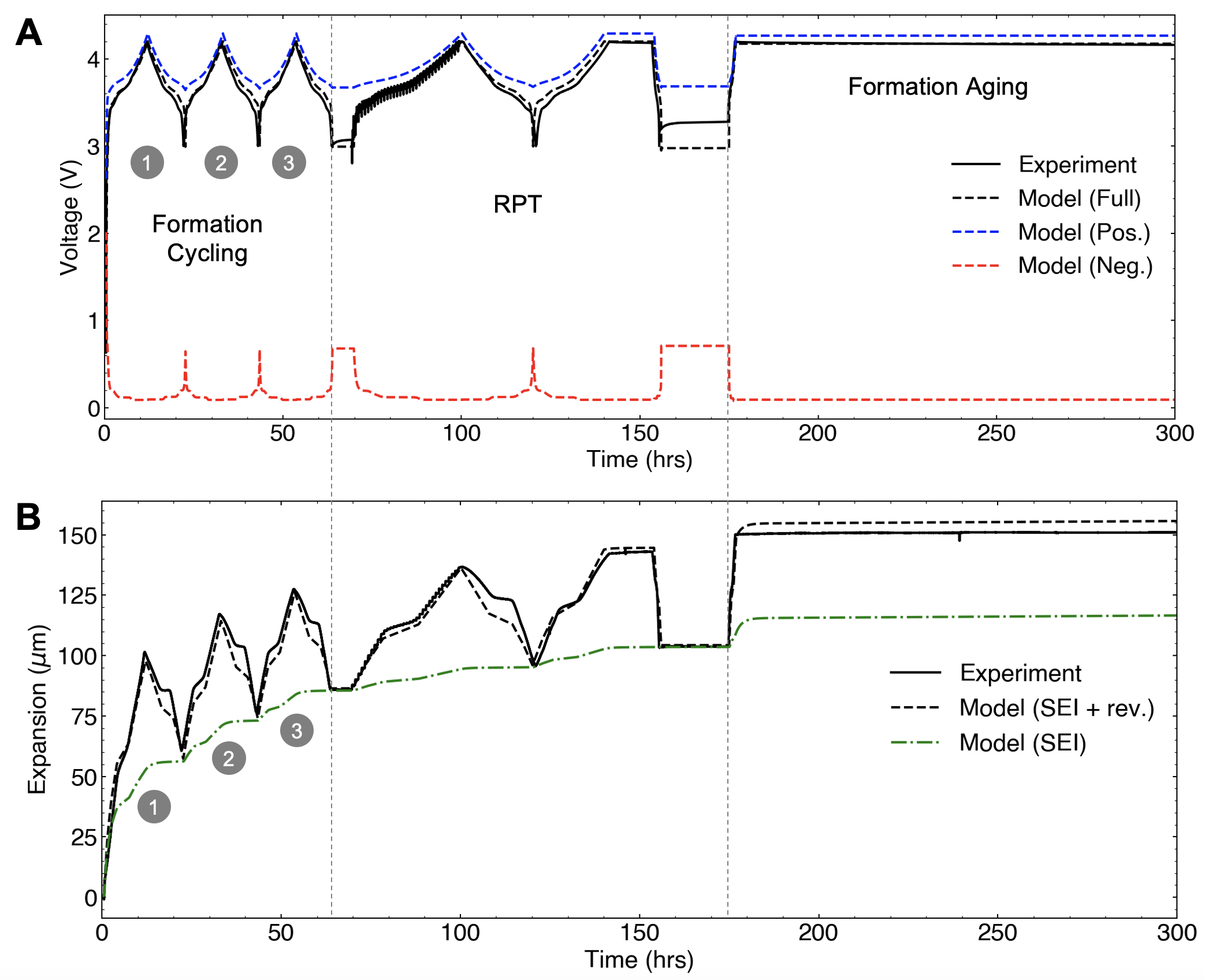}
\end{center}
\caption{\textbf{Model comparison against experimental data.} The dataset consists of three formation charge-discharge cycles, a reference performance test (RPT) sequence, and formation aging at 100\% SOC. (A) Full cell voltage. (B) Thickness expansion.}
\label{fig:model-validation}
\end{figure}

Figure \ref{fig:model-validation}A compares the modeled versus measured full cell voltage during and after the formation cycles. Model-predicted positive and negative electrode potentials vs Li/Li$^+$ are also shown for reference. The simulation was run using the parameters described in Section \ref{sec:parameters} and with a current input that matches the experimental data. The result shows that the formation model outputs generally match the observed voltage trends over the course of the formation charge-discharge cycles and during the subsequent formation aging step.

\rev{However, despite the general agreement between model and experiment, several model mismatches can be observed.} First, the cell voltage was under-predicted during the three formation discharge steps. This may be due to the fact that the half-cell equilibrium potential functions \citep{Mohtat2020-zp} were based on charge curves and thus do not account for hysteresis effects \citep{Birkl2015-no, Dreyer2010-il, Hulikal2017-zj}. The half-cell equilibrium potential functions were also inherited from a previous work \citep{Mohtat2020-zp} which had similar but not identical positive and negative electrode compositions. Next, the voltage rebound at 0\% SOC was not predicted by the model. This model mismatch is attributed to the fact that our overpotential model uses a cell resistance does not vary with SOC, yet \rev{it is well-known that cell resistance increases significantly at lower SOCs owing to kinetic} limitations in the NMC positive electrode \citep{Weng2021-qc, Yang2012-sp, Zhou2019-gw, Liu2021-rc}.

\subsection{Full Cell Expansion}

Figure \ref{fig:model-validation}B compares the modeled versus measured full-cell thickness expansion during and after the formation cycles. The model-predicted irreversible expansion due to SEI growth is also shown as a green dashed line. The model captures macroscopic trends in both irreversible and reversible full-cell expansion, across formation cycles and subsequent formation calendar aging. Consistent with the experiment, the model correctly predicts that the irreversible expansion slows down during the formation aging step. This result was achieved by using the boosted SEI growth formulation which accelerates the SEI growth process during cycling (see Section \ref{sec:boost}).

As with the voltage data, some model mismatches persist. First, the model over-predicts the electrode contraction at mid-SOCs during discharging steps. The origin of this mismatch is unclear, but we hypothesize that the graphite expansion function may also be subject to charge-discharge asymmetry. In this work, the expansion functions were parameterized on charge, not discharge \citep{Mohtat2020-zp}, so the model would not be able to capture this hysteresis effect. Another model mismatch occurs during the formation aging step. While the model correctly captures the slope of the expansion data, the model `over-boosted' the SEI growth rate, and thus the total expansion, during the C/2 charge leading up to the formation aging step. This result suggests that the first-order representation of the boosting mechanism given by Eq. \ref{eq:boost2} may be insufficient to capture the dynamics of boosting at all C-rates. \rev{Thus, while the SEI growth boosting mechanism represented in this work remains a reasonable starting point to capture expansion differences between cycling and calendar again, room remains to improve the representation of higher-order dynamical effects such as cycling C-rates.}

We finally note that active material loss mechanisms are not currently represented in the model, which could contribute to another source of model error. As the electrodes lose active material, (e.g. due to loss of electronic conduction pathways, or due to material phase transformations), the lost active material will no longer participate in reversible expansion, decreasing the overall observed reversible expansion. Finally, similar to the equilibrium potential functions, the volumetric expansion functions $\nun$ and $\nup$ are also inherited from literature \citep{Mohtat2020-zp} and may not be exact, contributing to yet another possible source for model error.

\subsection{Formation Coulombic Efficiencies}

\begin{figure}[ht!]
\begin{center}
\includegraphics[width=0.7\linewidth]{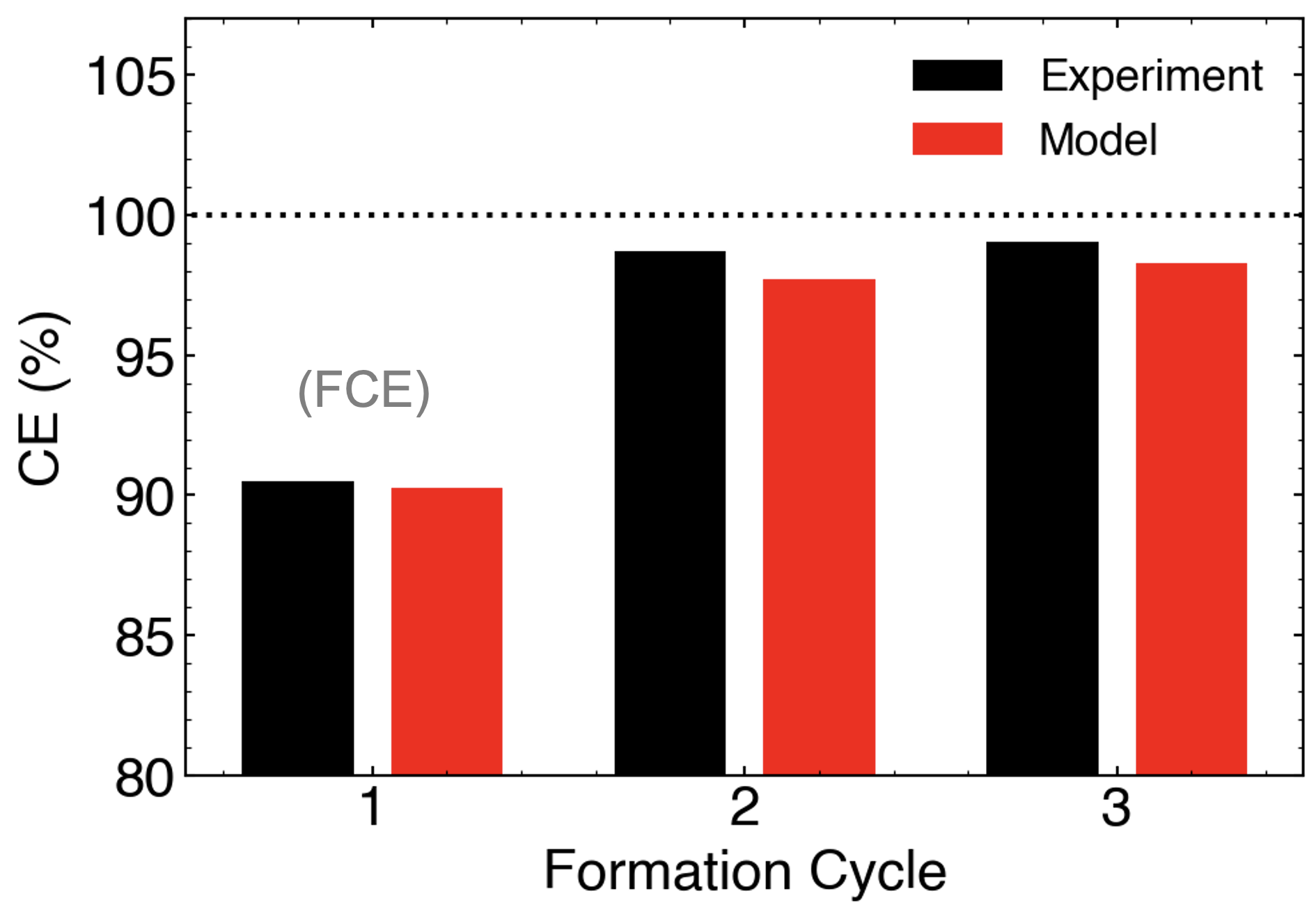}
\end{center}
\caption{\textbf{Model vs measured Coulombic efficiency (CE).} First cycle efficiency (FCE) is defined as the CE of the first formation charge cycle.}
\label{fig:model-ce}
\end{figure}

Figure \ref{fig:model-ce} compares the modeled versus measured FCE during the first formation cycle as well as the CE of the subsequent two cycles, defined by Eq. \ref{eq:ce}. The formation model predicted the FCE within 1 percentage point of the measured result. However, the model underestimated the CE of the next two cycles. The model mismatches here are attributed to suboptimal parameter tuning which can be improved \rev{with further optimization.}

Overall, the formation model not only predicted that the FCE is much lower than the CE of subsequent cycles, but the model also gave an intuitive explanation for why this is necessarily the case: the first charge cycle is the only cycle that sees reaction-limited SEI growth. After the first charge cycle, the SEI reaction becomes diffusion-limited, slowing down the SEI reaction rate for all subsequent cycles (see Section \ref{sec:reaction-diffusion}).

\subsection{Formation Charge dQ/dV}

\begin{figure}[ht!]
\begin{center}
\includegraphics[width=1.00\linewidth]{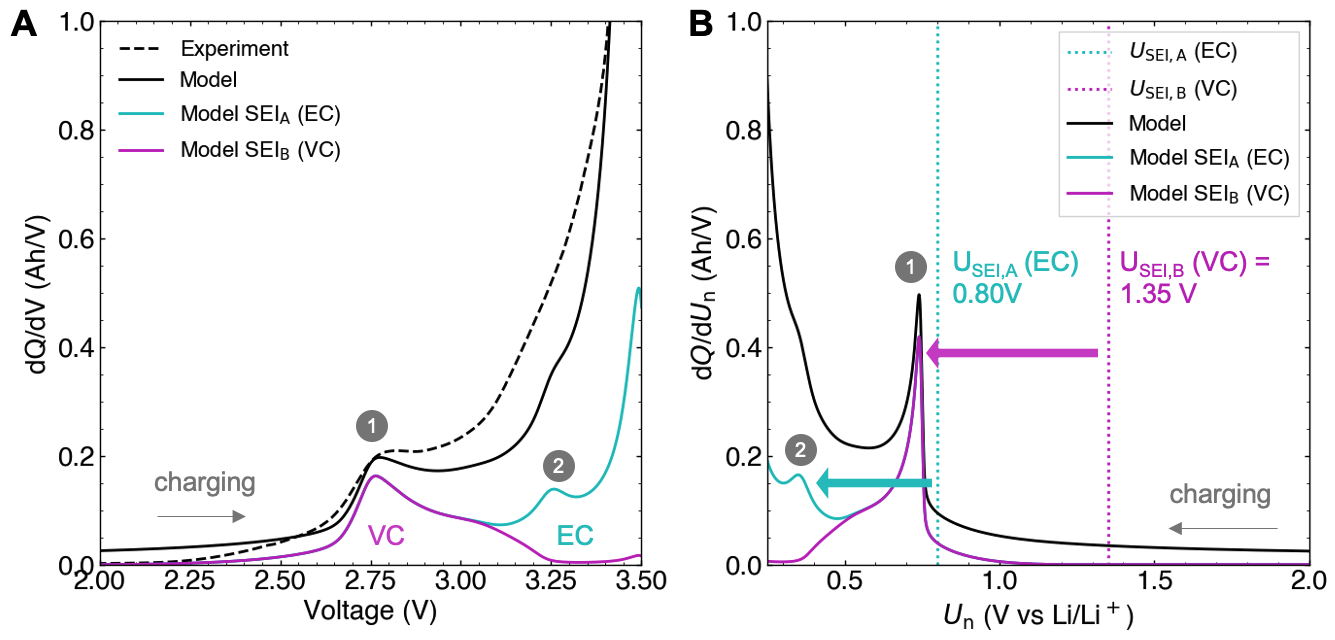}
\end{center}
\caption{\textbf{Modeled vs experimental d$Q$/d$V$ trace during the formation charge cycle.} (A) d$Q$/d$V$ vs $V$, where $V$ is the full cell voltage. (B) d$Q$/d$\Un$ vs $\Un$, where $\Un$ is the negative electrode equilibrium potential.}
\label{fig:dqdv}
\end{figure}

Figure \ref{fig:dqdv}A compares the model vs measured full cell dQ/dV curve during the first formation charge. The plot represents the incremental capacity accumulated by the cell at each increment of full cell voltage. This plot is experimentally observable, with the measured result plotted as a dashed line. This plot is commonly studied in electrolyte development studies since peaks in dQ/dV during the initial formation charge cycle can often be attributed to SEI-forming electrolyte reduction processes. However, care is usually needed in interpreting the data since the dQ/dV peaks can also be due to shifts in the solid-phase potentials as the positive electrode is delithiated and the negative electrode is lithiated. 

In the experimental data, a dQ/dV peak was observed at approx. 2.8V (marker 1). The formation model attributed this peak to the reduction of the VC additive. The model predicted the position of this reduction peak and also qualitatively captured the peak shape. The model additionally predicted a second, broad peak, corresponding to EC reduction (marker 2). However, the EC peak was not observed in the experimental data. \rev{This model mismatch may be due to differences in the electrolyte composition: the experimental data was obtained with an electrolyte comprising both EC and DEC, but the model only represented EC.}

\section{Discussion: Mechanistic Insights}
\label{sec:discussion}

The formation model developed in this work provides a quantitative understanding of what makes the first formation charge process different from all subsequent cycles in lithium-ion battery systems. This understanding holds important practical implications as well as mechanistic insights, which we discuss in this section. 

\subsection{What is Special About the First Formation Charge?}

\textit{Minimal SEI growth prior to formation charge.} The formation model predicts that the SEI growth does not begin until the cell externally charged for the first time. Before the charge step, the SEI formation process is suppressed since the negative electrode potential, at ca. 2.0V, is above the reaction potentials for the reacting molecules studied for this work (0.8V for EC and 1.35V for VC), \rev{suppressing the reaction currents according} to Eq. \ref{eq:jsei-bv}). This observation has some practical implications: since SEI formation only begins when the cell is charged, the duration elapsed between electrolyte filling and the first formation charge cycle may be discounted as a potential source of variability in SEI properties and subsequent cell performance. Note, however, that copper dissolution at high negative electrode potentials, which may occur during the time between electrolyte filling and formation charge \citep{Mao2018-vv, Hendricks2020-xr}, creates a separate source of variability that warrants further study using the formation model.

\textit{Reaction-limited SEI growth occurs exclusively during formation charge.} The first cycle efficiency (FCE) during formation is typically between 80\% to 90\% and can vary as a function of the electrolyte system \citep{Eldesoky2021-fx}, negative electrode type \citep{Eldesoky2021-fx}, and formation protocol \citep{Mao2018-vv}. The FCE is thus much lower than subsequent measures of CE which typically exceeds 99\%. This result was experimentally replicated in our formation dataset and was also predicted by our formation model, see Figure \ref{fig:model-ce}. The formation model provides a simple explanation to why the FCE is low compared to the CE of subsequent cycles: the first formation charge is the only time the SEI growth process is reaction-limited. After the first charge cycle, SEI growth becomes diffusion-limited (see Figure \ref{fig:model-diffusion-reaction}). Intuitively, the SEI film is the thinnest during the first formation charge, making solvent diffusion facile. During the first formation charge, a sufficient amount of SEI is built up and the SEI reaction becomes self-limiting due to diffusion limitations. For the remaining life of the battery, the SEI reaction remains diffusion-limited. This result suggests that the effective SEI diffusivity, $\Dsei$, is a key parameter describing the ``passivation'' of an SEI. A low effective SEI diffusivity is thus essential for creating a long-lasting battery. \rev{By contrast, the SEI kinetic rate parameter, $\ksei$, appears to mostly play a role in determining the SEI formation dynamics during the first charge cycle. The model predicts that $\ksei$ becomes inconsequential to the SEI growth rate during later formation cycles and for the rest of the cell's life.}

\begin{figure}[ht!]
\begin{center}
\includegraphics[width=1.0\linewidth]{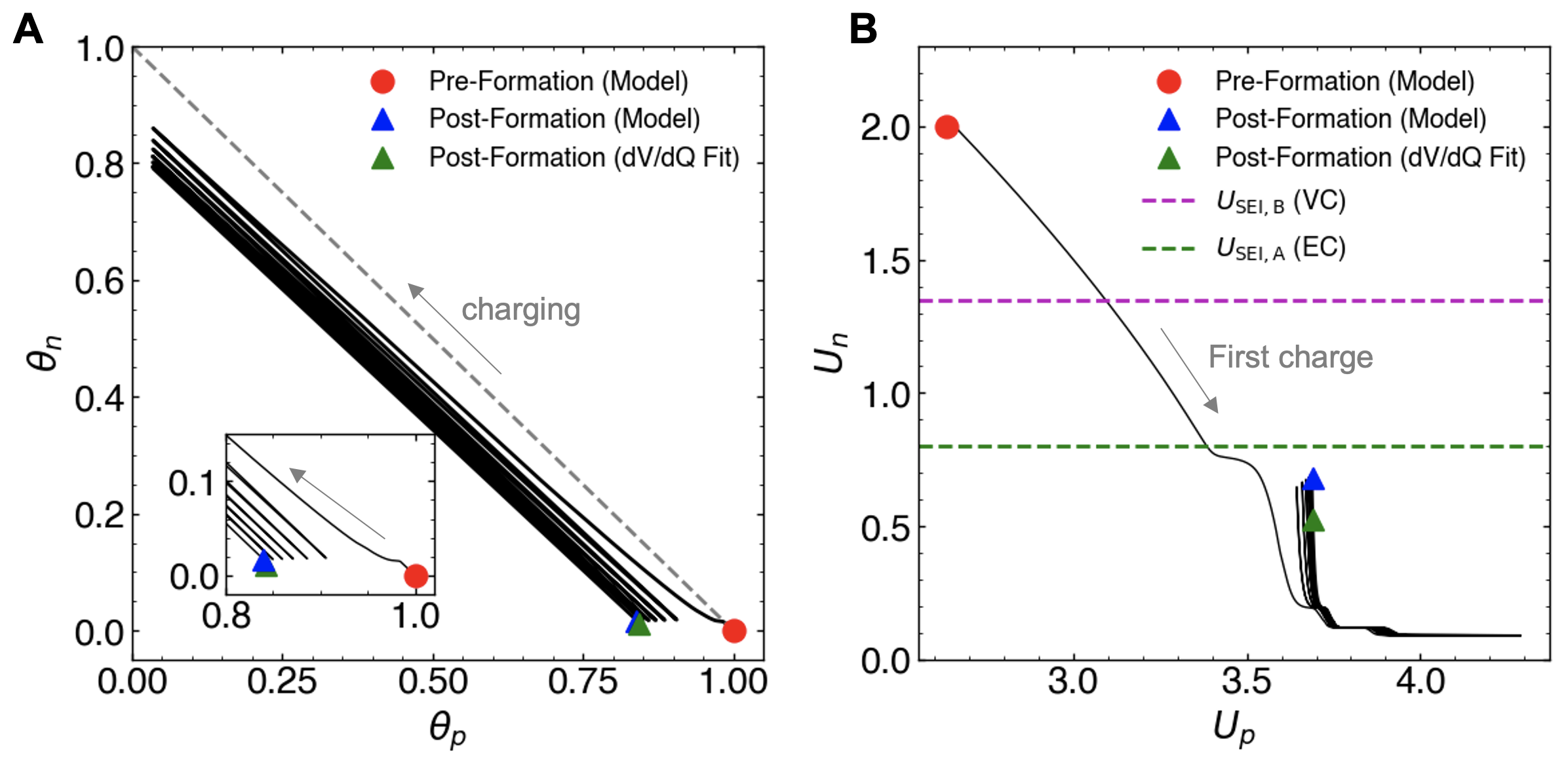}
\end{center}
\caption{\textbf{Lithium stoichiometry and electrode potential maps over formation cycling, RPT, and formation aging.} (A) Positive and negative electrode lithium stoichiometries. Inset: expanded view of the lithium stoichiometries at the end of each discharge cycle. (B) Positive and negative electrode equilibrium potentials. Red circle: state of the system before formation. Triangles: state of the system after formation completes and the cell is discharged. Green triangle: prediction from dV/dQ fitting method on C/20 charge voltage data taken at the end of the formation process, adapated from Weng et al. \citep{Weng2023-lj}.}
\label{fig:model-stoichiometry}
\end{figure}

\textit{Extended lithium stoichiometric ranges and electrode surface potentials.} Figure \ref{fig:model-stoichiometry} shows how the lithium stoichiometries in each electrode evolve during and after the formation cycles (Panel A), along with their corresponding electrode potentials (Panel B). Before the first formation charge, the negative electrode is completely empty of lithium ($\thetan=0$) and the positive electrode is completely full of lithium ($\thetap=1)$, shown in red circles. After formation completes, neither electrode will be able to access these initial lithium stoichiometries and their corresponding electrode potentials. Notably, the maximum lithium stoichiometry in the positive electrode continues to decrease over the RPT and formation aging (Panel A inset), a result directly due to the irreversible consumption of lithium from the positive electrode to form the SEI. By the end of the formation aging step, the maximum positive electrode stoichiometry has decreased from 1.0 to ca. 0.8 (blue triangles). Meanwhile, the minimum negative electrode potential has also increased from 0.0 to ca. 0.018. In both electrodes, these permanent shifts in the accessible lithium stoichiometry range also restricts the range of electrode potentials observable after formation completes \rev{(Panel B)}. The negative electrode starts at 2.0V vs Li/Li$^+$ before formation, but by the time the first formation charge completes, the negative electrode will never see potentials above 0.7V vs Li/Li$^+$. These restrictions in electrode potentials may not bear much consequence to the overall SEI reaction rate over life, since the SEI reactions will be primarily driven by diffusion limitations as explained in Section \ref{sec:reaction-diffusion}. However, this result does explain why copper dissolution, which occurs at high negative electrode potentials, is a concern during the formation process, but is less of a concern over the remainder of a battery's life.

\subsection{Reinterpreting dQ/dV Peaks for SEI Reaction Studies} 

Returning to Figure \ref{fig:dqdv}, Panel B shows a dQ/dV plot where the differential capacity is plotted against the differential negative electrode equilibrium potential $\Un$. This result is discussed briefly now to gain further insight into the rate-limiting mechanisms during formation charge. A key observation here is that the position of the reduction peaks for both EC and VC do not indicate their respective reaction potentials, as is often assumed in previous literature studies. Rather, there is a delay between the reaction potentials and the respective reduction peaks. For example, the VC reaction potential is at 1.35V, yet the observed dQ/dV peak does not occur until the negative electrode has dropped below 0.8V. Similarly, the EC reaction potential is at 0.8V, but its corresponding dQ/dv peak occurs below 0.4V. This delay suggests that the dQ/dV peaks do not indicate the reaction potentials; rather, the peak indicates the transition from the reaction-limited to the diffusion-limited regime, which occurs sometime after the negative electrode potential falls below the reaction potential (see Figure \ref{fig:model-diffusion-reaction}. This result partly explains why literature reports of SEI reaction potentials, $\Usei$, can vary widely. 

\subsection{Analyzing Rate-Liming Regimes} 

The transition from reaction-limited SEI growth to diffusion-limited SEI growth appears to be a key characteristic of the first formation charge cycle. Due to its importance, we briefly discuss some analytical approaches to further characterizing this transition to promote future investigations in this area. 

We define a dimensionless number, $\Bu$, as:
\begin{equation} \label{eq:mx}
    \Bu \triangleq \frac{\jseirxn}{\jseidif},
\end{equation}
$\Bu$ characterizes the ratio of SEI current carried by the reaction versus the diffusive process. When $\Bu \ll 1$, the SEI current is reaction-limited, and when $\Bu \gg 1$, the SEI current is diffusion-limited. $\Bu$ can be found by substituting Eqs. \ref{eq:jsei-rxnmax} and \ref{eq:jsei-difmax} into Eq. \ref{eq:mx}, yielding:
\begin{equation}
    \Bu = \frac{\krxn}{\kdif},
\end{equation}
where:
\begin{align}
    \krxn &\triangleq \ksei\exp\left(-\frac{\alphasei F}{RT}\etasei\right) \\
    \kdif &\triangleq \frac{\Dsei}{\deltasei}.
\end{align}
$\krxn$ is the apparent reaction-limited rate parameter which depends on both the pure rate constant $\ksei$ as well as the reaction over-potential $\etasei = f(\Un,\Usei)$. $\kdif$ is the apparent diffusion-limited rate parameter which accounts for both the effective diffusivity as well as the film thickness. These rate parameters clarify that the rate-limiting process depends on both internal material properties, such as $\ksei$ and $\Dsei$, as well as external factors such as $\Un$ and $\deltasei$.

\subsection{Implications of SEI Growth Boosting}

The boosted SEI growth mechanism and its mathematical representation (see Sections \ref{sec:boost}, \ref{sec:boost-explore}) were key developments enabling the model to capture experimental trends during both formation cycling and formation aging. Here, we further clarify the modeling assumptions and physical implications of the SEI boosting mechanism, highlighting its role in enabling the formation model to be extended towards general-purpose cycle life and calendar aging simulations.

\textit{Necessity.} A key insight from this work was that a boosted SEI growth mechanism during charging was necessary to capture the macroscopic trends in capacity fade and cell expansion during formation cycling and formation aging. Without the boosted SEI mechanism, the model could only fit formation cycling and aging using separate parameter sets, but not both using the same parameter set. Specifically, without boosting, SEI diffusivities that explained the voltage and expansion data during formation cycling led to unrealistically high degradation rates during formation aging. Conversely, an SEI diffusivities that explained the voltage and expansion data during formation aging was too small to explain the cell expansion trend during formation cycling. However, when the boosted SEI growth mechanism was introduced, both formation cycling and aging was explained by the same set of SEI diffusivities (see Figure \ref{fig:model-validation}). Introducing the boosted SEI mechanism thus enabled a unified representation of SEI growth rates across cycling and storage. 

\textit{Reaction rate asymmetry.} The boosted SEI growth mechanism also introduces a source of reaction rate asymmetry: SEI growth is predicted to occur more quickly during charging (negative electrode lithiation) compared to discharging (negative electrode delithiation). This result is consistent with experimental results from Attia and Das \citep{Attia2019-jg, Das2019-la}. The boosting mechanism frames the reaction rate asymmetry as a particle stress-driven phenomenon whereby reaction rates are boosted only when the SEI expansion but not when it shrinks. Boosting thus provides a unified description of the apparent SEI reaction rate asymmetry described in earlier works. 

\textit{Capacity fade does not trend with square root of time}. Attia et al. \citep{Attia2020-hn} reported that experimental data on lithium-ion battery degradation often did not strictly follow a $\sqrt{t}$ relationship, as would be expected with diffusion-limited SEI growth. The boosted SEI growth mechanism provides a quantitative description of why the $\sqrt{t}$ relationship is violated, and more specifically, why the exponent is often more than 0.5. During charging, the boosting mechanism enhances the SEI reaction rate, taking the system out of any steady-state diffusion regime it may have been previously. The $\sqrt{t}$ dependence can thus be viewed as a limiting case where the boost factor $\gamma$ is zero.

\textit{Model representation.} In our model, the SEI growth boosting process was represented as a modification to the effective diffusivity $\Dsei$. In reality, the effective reaction rate parameter $\ksei$ could also be boosted since the reaction surface area could increase during SEI cracking. However, in the model, increasing $\ksei$ will have little impact on the SEI growth rate beyond the initial formation charge cycle since the SEI growth process becomes diffusion-limited by the end of the first cycle (see Figure \ref{fig:model-diffusion-reaction}). Overall, our treatment of SEI growth boosting remains phenomenological: the model captures the macroscopic trends in voltage and expansion without elucidating the mechanistic details of the SEI boosting process. Such mechanistic details would be more appropriate to address using higher-fidelity models.

\section{Future Work and Applications}

\reva{This paper focused on developing the mathematical preliminaries necessary for building practical models of the battery formation process. Towards this end, we adopted a `dual-tank' model to contextualize the SEI formation process within a full cell system. We believe that this level of representation captures enough physically-relevant processes to yield useful predictions while being simple enough to make parameter-tuning feasible. Ultimately, the SEI growth dynamics predicted by the model bear measurable consequences to the initial cell state, including the FCE, the initial cell capacity, and long-term degradation trajectories. We thus envision that this model can be deployed towards several applications in cell manufacturing, including formation protocol optimization, electrolyte design, cell lifetime prediction, and degradation model parameter identification. This section describes these future applications and identifies relevant next steps to realize these applications in practice.}

\subsection{SEI Passivation and SEI Diffusivity as Cell Lifetime Indicators}
\label{sec:passivation}

\reva{Perhaps the most important role of battery formation is to create a so-called `passivating SEI.' Such an SEI would minimize further SEI reactions after the formation process completes, leading to a longer-lasting cell. A more passivating SEI is typically considered to be more dense or less porous} \citep{Attia2021-bl}. Conversely, a less passivating, or `non-passivating' SEI, is assumed to be less dense or more porous.

The formation model represents the concept of SEI passivation through the homogenized effective SEI diffusivity, $\Dseirbar$, see Eq. \ref{eq:deff}. This parameter encapsulates the effect of SEI porosity according to Eq. \ref{eq:dseieff}: a less porous, more passivating SEI will have a lower $\Dseirbar$. Our model therefore interprets the formation of a passivating SEI as the formation of an SEI that minimizes $\Dseirbar$. Note that the SEI reaction rate parameter, $\ksei$, does not play a role in determining SEI passivation since the system becomes diffusion-limited at the end of the very first formation charge cycle (see Section \ref{sec:reaction-diffusion}).

The formation model eludicates how SEI passivation can be tuned: $\Dseirbar$ depends on both the electrolyte composition and the formation protocol. The electrolyte composition sets the individual solvent diffusivities, $\Dseirl$, which are material properties that may be tunable through electrolyte engineering. Meanwhile, the formation protocol can be used to tune how much time is spent at each negative electrode potential. In theory, the formation protocol can be designed to preferentially reduce one particular electrolyte component associated with a low diffusivity, resulting in an overall more passivating SEI film without changing the electrolyte composition.

To summarize, the formation model can be used to predict how electrolyte composition and formation protocols can influence SEI passivation, which ultimately sets the long-term SEI growth rate and hence the cell lifetime. Future work in this area should focus on exploring the sensitivity of electrolyte composition and formation protocols on the resulting SEI diffusivities.

\subsection{Direct Lifetime Simulations and Path-Dependent Aging}

\reva{In addition to predicting cell lifetime indicators such as SEI passivation, we envision that the formation model can be used to directly simulate cell lifetime. This can be done by extending the simulation time domain to include cycling and calendar aging conditions after the formation process completes. The simulation can, in theory, be run until the cell capacity drops below some minimum threshold, which could encompass hundreds or thousands of cycles with varying usage profiles. We expect that such usage of the formation model is valid since SEI growth is generally understood to be the dominant degradation mechanism in many commercial lithium-ion systems and under typical use cases }\citep{Preger2020-ln, Weng2021-qc}. 

\reva{Using the formation model to directly simulate aging confers several advantages. First, the formation model provides a consistent representation of an essential degradation process - SEI growth - during the formation process and over the remainder of a cell's life. The resulting simulation therefore bridges the gap between battery manufacturing process understanding and resulting cell lifetime performance. Second, the formation model captures path-dependent aging through the boosted SEI growth mechanism. With this mechanism, the boost sensitivity factor, $\gamma$, from Eq. }\ref{eq:boost2} can be viewed as a `cycling loss sensitivity' factor. As $\gamma$ approaches zero, no SEI growth boosting occurs during cycling and the diffusion process remains under steady-state. Under this condition, the cell experiences only calendar aging losses. Such a scenario could describe `power cells' having high electrode porosities and low energy densities. Meanwhile, as $\gamma$ increases, the SEI growth process during cycling becomes boosted, so the cell becomes more sensitive to damage during cycling. Such a scenario could describe `energy cells' having low electrode porosities and high energy densities. With the boosting mechanism, dynamical changes to the aging profile, including transitions between cycling and aging conditions, are naturally represented by the model. Finally, the formation model, being a zero-dimensional model, is fast to run, making it suitable for lifetime simulations which often involve simulating hundreds to thousands of cycles.

The formation model simulation framework already allows for the simulation of arbitrary cycling and calendar aging profiles. The focus of next steps should therefore focus on validating simulation results against experimental lifetime data. Gaps identified between modeled versus measured results could motivate further model developments, which could include refining the boosted SEI growth mechanism representation, \reva{including other degradation modes beyond SEI growth such as active material losses}, and improving the numerical implementation.

\subsection{Formation Protocol Optimization and Electrolyte Design}

Since the formation process is time and energy-intensive, battery manufacturers are incentivized to decrease the time taken for the formation and aging process \cite{Liu2021-ye, Wood2019-uu}. But how does decreasing formation time affect SEI passivation? The formation model clarified how SEI passivation during formation depends on both electrolyte properties (e.g. electrolyte composition, additive amount, diffusivities, reaction potentials, rate constants) and the formation protocol (e.g. time spent at different negative electrode surface potentials, the negative electrode expansion rate). Identifying which of these parameters more strongly influence cell characteristics (e.g. FCE, initial cell capacity, long-term degradation rates) can help inform both electrolyte design and formation protocol design. One particular question that may be answered is: ``how much influence does the formation protocol have on determining initial cell characteristics?'' 

Future work to evaluate the feasibility of model-based formation protocol optimization should focus on validating model outputs against `fast formation' protocols already proposed in literature \citep{Weng2021-qc, An2017-qy}. Similarly, predictions made for different electrolyte compositions and properties will also need validation against experiments.

\subsection{Degradation Model Parameter Identification}

Physics-based battery degradation models are typically tuned, or calibrated, against cycling and calendar aging data. See, for example, work by Reniers et al. \cite{Reniers2023-bg}. These experimental datasets are collected on cells after formation has already completed. Our work introduces an opportunity to leverage a new dataset - formation cycling and aging - to improve model parameter identification. The degradation parameters from the formation model (e.g. $\Usei$, $\Dsei$, $\ksei$) may be directly transferable to other modeling contexts such as the Single Particle Model \cite{Guo2010-av, Moura2017-lg} and the full Doyle-Fuller-Newman (or P2D) model \cite{Fuller1994-pm, Chen2022-zc}. The experimental formation data provides additional, physically-relevant features, such as voltage traces during formation cycling and aging (Figure \ref{fig:model}), the FCE (Figure \ref{fig:model-ce}) and the first charge dQ/dV (Figure \ref{fig:model-stoichiometry}), which all serve to constraint the set of possible degradation model parameters. Additionally, macroscopic expansion measurements can also be used to tune electrode parameters such as particle sizes and electrode dimensions (Eq. \ref{eq:expansion}) as well as SEI properties such as SEI molar volume (Eq. \ref{eq:ddeltadt}).

An expected obstacle for model parameter identification relates to the problem of parameter uniqueness, or parameter identifiability. Even with our reduced-order SEI growth representation, each electrolyte component is described by at least 4 intensive material properties: $\Dsei$, $\Usei$, $\ksei$, and $\vmsei$. While the presence of expansion measurements is expected to improve the identifiability of some parameters, it may still not be enough to identify all of the parameters. More ex-situ characterization may thus be generally needed to reduce the number of model parameters that need tuning. A suggested next step for this work is to first understand model parameter sensitivity, to answer the question: ``which model parameters have the greatest impact on observed cell characteristics such as FCE, initial capacity, and degradation rates?''

\section{Conclusions}

This work develops a reduced-order electrochemical model of the formation cycling and aging process for lithium-ion batteries. The model supports simulating an arbitrary number of electrolyte components that react at the negative electrode surface to form a composite SEI film with volume-averaged properties. The model also tracks electrode-level lithium stoichiometries and potentials, enabling the tracking of SEI current densities as a function of full-cell voltages. The model further couples an electrode expansion model which accounts for both reversible and irreversible expansion. We finally introduce an SEI growth boosting mechanism which enabled a unified description of SEI growth during both formation cycling and formation calendar aging. We demonstrated that the formation model qualitatively matched experimental formation data, including cell voltages, cell expansion, coulombic efficiencies (including first cycle efficiency, or FCE), and the dQ/dV curve during the first charge cycle which indicated the reduction of the VC electrolyte additive. 

The formation model clarified why the first formation charge cycle is special. First, SEI growth does not begin until external current is applied to the cell. Second, the SEI growth process is reaction-limited during the first formation charge but becomes diffusion-limited for the remaining life of the cell. Third, during the first formation charge, each electrode sees an extended range of lithium stoichiometries and electrode surface potentials which will never be accessed again for the remaining life of the cell.

We envision that the formation model can be leveraged for improving formation protocol design, electrolyte engineering, cell lifetime simulations, and physics model parameter identification. In a battery manufacturing context, our model could see potential applications as a physics-based digital twin of the formation process. 

Overall, the proposed modeling framework enables practical pathways for bridging the electrochemistry of battery formation to macroscopic variables related to battery performance, safety and lifetime.

\section*{Nomenclature}

\noindent Acronyms
\begin{itemize}
    \item CC: constant current
    \item CE: coulombic efficiency
    \item CV: constant voltage
    \item DEC: diethyl carbonate
    \item DMC: dimethyl carbonate
    \item EC: ethylene carbonate
    \item FCE: first cycle efficiency
    \item $\liedc$: \rev{lithium ethylene dicarbonate}
    \item LVDC: lithium vinylene dicarbonate
    \item SEI: solid electrolyte interphase
    \item VC: vinylene carbonate
    \item SOC: state of charge
\end{itemize}

\noindent Subscripts
\begin{itemize}
    \item $i$ : electrode ($\mathrm{n}$ or $\mathrm{p}$)
    \item $r$ : SEI reacting species (A, B, ...)
    \item $l$ : SEI solid reaction product (A', B', ...)
\end{itemize}

\noindent Superscripts
\begin{itemize}
    \item $o$ : electrolyte bulk phase
\end{itemize}

\begin{appendices}

\begin{figure}[ht!]
\begin{center}
\includegraphics[width=1.0\linewidth]{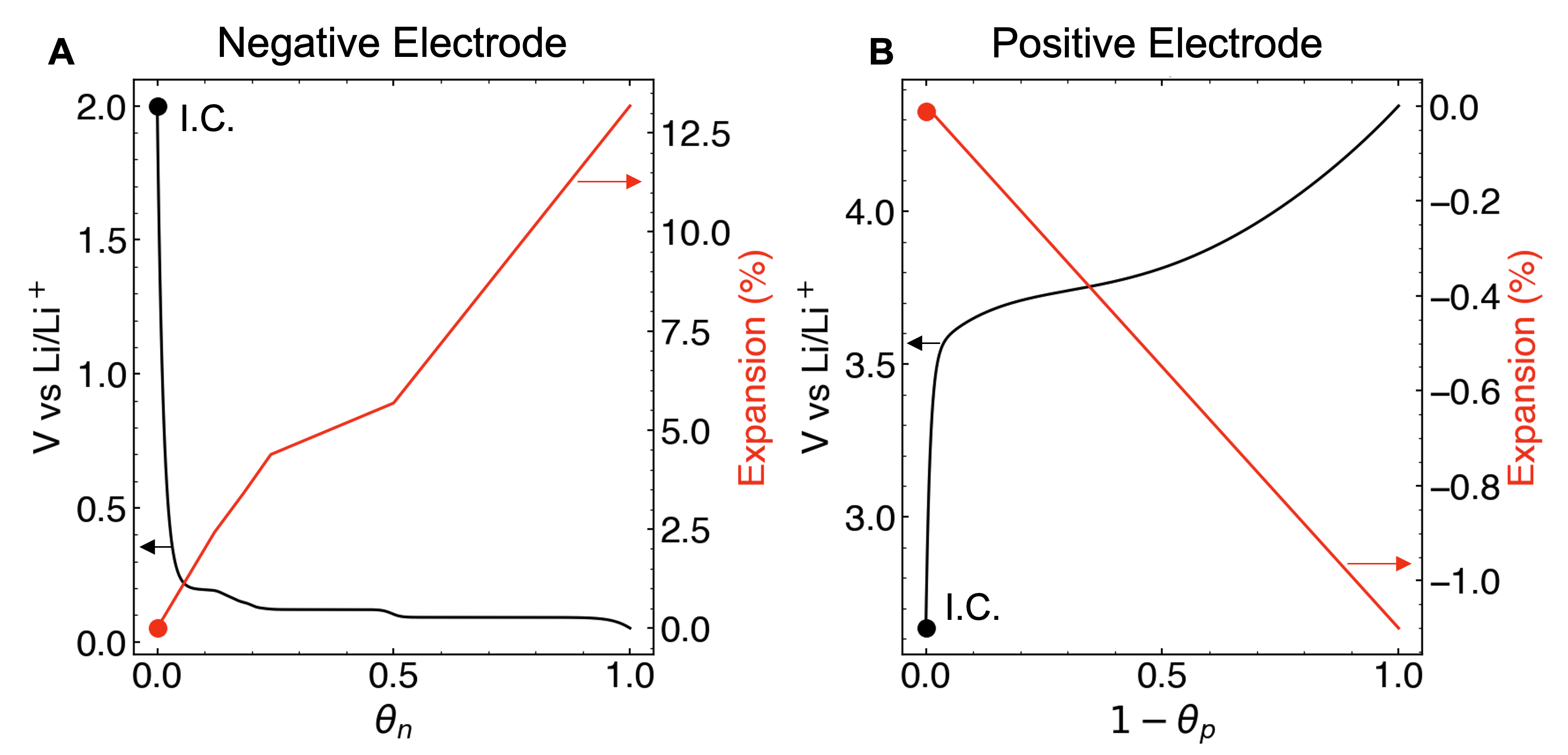}
\end{center}
\caption{\textbf{Half-cell near-equilibrium potential functions and volumetric expansion functions.} These functions were adapted from Mohtat et al. \citep{Mohtat2020-zp}. The potential functions have additionally been expanded to satisfy Eq. \ref{eq:voltage-bc}. (A) Negative electrode (graphite) $\Un$ and $\nun$. (B) Positive electrode (NMC622) $\Up$ and $\nup$.}
\label{fig:ocp}
\end{figure}

\section{Expansion Model Derivations} \label{sec:expansion-derivation}

\subsection{Reversible Expansion from Intercalation-Induced Electrode Swelling}

Unit cell lattice parameters can be converted into a net volume change $\nu_i$ according to: 
\begin{equation} \label{eq:deltai}
    \nu_i = \frac{dV_i}{V_i},
\end{equation}
where $V_i$ is the volume of a unit cell. To relate the experimental data $\nu_i$ to the macroscopically observed cell thickness change, we first define the volume-averaged number of particles along the total length of the electrodes, $\bar{N}_p$, as:
\begin{equation} \label{eq:nu1}
    \bar{N}_p \triangleq \frac{L_i}{R_i}
\end{equation}
where $\Li$ is the electrode thickness and $\Ri$ is the radius of a single electrode particle which is assumed to be spherical. We assume that $\bar{N}_p$ remains constant as the particles expand and contract, and hence:
\begin{equation} \label{eq:nu2}
    \bar{N}_p = \frac{dL_i}{dR_i},
\end{equation}
where $d\Li$ is the incremental thickness change of the electrode and $d\Ri$ is the corresponding electrode particle radius change. Combining Eqs \ref{eq:nu1} and \ref{eq:nu2} yields a scaling law to convert from microscopic expansion to macroscopic expansion.
\begin{equation} \label{eq:scaling}
    d\Li = \frac{L_i}{R_i}dR_i.
\end{equation}
Next, we convert $\nu_i$ to radial expansion, $dR_i$, by assuming a spherical particle, in which case:
\begin{equation} \label{eq:nu-to-dr}
    dR_i = \frac{R_i}{3}\nu_i,
\end{equation}
where we have substituted the equation of a sphere for particle volume $V_i$, with the differential form $dV_i = 4\pi R_i^2dR_i$. We can now substitute Eq. \ref{eq:nu-to-dr} into Eq. \ref{eq:scaling} to arrive at:
\begin{equation}
    dL_i \triangleq = \frac{L_i}{3}\nu_i.
\end{equation}
Finally, the total reversible cell thickness expansion from electrode $i$ is:
\begin{align}
    \Delta_i &= N_\mathrm{layers} \cdot dL_i \\
             &= \frac{N_\mathrm{layers}L_i}{3}\nu_i,
\end{align}
where $N_\mathrm{layers}$ is the number of electrode layers in the stacked pouch cell configuration.

\subsection{Irreversible Expansion from SEI Growth}

Assuming constant particle density, we have that:
\begin{equation}
\rhosei=d\msei/d\vsei,
\end{equation}
where $\msei$ is the SEI mass accumulated in kg and $\rhosei$ is the SEI density in kg/m$^3$, and $\vsei$ is the SEI volume in m$^3$. SEI mass accumulation is proportional to total charge accumulated, $\Qsei$, according to: 
\begin{equation} \label{eq:mass-charge}
d\msei =  \frac{\Msei}{n F}d\Qsei,
\end{equation}
where $\Msei$ is the molar mass and $\Qsei=\int \Isei dt$ is the charge accumulated to form SEI. It follows that:
\begin{equation} \label{eq:volume-charge}
d\vsei = \vmsei\frac{d\Qsei}{n F}.
\end{equation}
We can then use \ref{eq:mass-charge} and \ref{eq:volume-charge} to derive an expression for the SEI thickness on the negative electrode particles:
\begin{align}
     \deltasei &= \frac{d\vsei}{N\cdot4\pi \Rn^2} \\
               &= \frac{1}{N\cdot4\pi \Rn^2}\vmsei\frac{d\Qsei}{nF}
\end{align}
where $N$ is the number of negative electrode particles. Differentiating both sides and subsituting $\Isei=d\Qsei$/$dt$ and $\jsei=\Isei/(N\cdot 4\pi \Rn^2)$ yields the expression:
\begin{equation}
\frac{d\deltasei}{dt} =\vmsei\frac{\jsei}{nF}.
\end{equation}
Eq. \ref{eq:scaling} can then be used to convert the SEI thickness growth into a macroscopic expansion:
\begin{align}
    dL_\mathrm{SEI} &= \frac{\Ln}{\Rn + dR_\mathrm{n}}\deltasei \\
                    &= \frac{\Ln}{\Rn(1 + \nun/3)}\deltasei,
\end{align}
where we have assumed that the SEI grows on the negative electrode particle with a radius $\Rn + dR_\mathrm{n}$. The total irreversible cell thickness expansion is then:
\begin{align}
    \deltasei &= N_\mathrm{layers}\cdot dL_\mathrm{SEI} \\
              &= \frac{N_\mathrm{layers}\Ln}{\Rn(1+\nun/3)}\deltasei,
\end{align}
where $N_\mathrm{layers}$ is the number of electrode layers in the stacked pouch cell configuration.

\end{appendices}

\bibliographystyle{ieeetr}
\bibliography{main}  

\end{document}